\documentclass[a4paper,fleqn,usenatbib]{mnras}
\usepackage{amsmath}
\usepackage{amssymb}
\usepackage{graphicx}
\usepackage{grffile}
\usepackage{float}
\usepackage[dvips]{epsfig}
\usepackage{epsfig}
\usepackage{dblfloatfix}
\usepackage{color}
\usepackage{caption}
\usepackage{hyperref}
\usepackage{bm}
\usepackage[british]{babel}

\newcommand{\lya}{\ifmmode{{\rm Ly}\alpha}\else Ly$\alpha$\ \fi}
\newcommand{\kms}{\ifmmode\mathrm{km\ s}^{-1}\else km s$^{-1}$\fi}
\newcommand{\vrot}{\ifmmode v_{\mathrm{rot}}\else $v_{\mathrm{rot}}$~\fi}
\newcommand{\vout}{\ifmmode v_{\mathrm{out}}\else $v_{\mathrm{out}}$~\fi}
\newcommand{\tauh}{\ifmmode \tau_{\mathrm{H}}\else $\tau_{\mathrm{H}}$~\fi}
\newcommand{\vth}{\ifmmode v_{\mathrm{th}}\else $v_{\mathrm{th}}$~\fi}
\newcommand{\hatk}{\ifmmode \hat{k}\else $\hat{k}$~\fi}
\newcommand{\STD}{\ifmmode \mathrm{STD}\else $\mathrm{STD}$~\fi}
\newcommand{\SKW}{\ifmmode \mathrm{SKW}\else $\mathrm{SKW}$~\fi}
\newcommand{\BI}{\ifmmode \mathrm{BI}\else $\mathrm{BI}$~\fi}

\begin{document}

\title[Lyman-$\alpha$ photons through rotating
  outflows]{Lyman-$\alpha$ photons through rotating outflows}  
\author[M.C. Remolina-Gutierrez \& J.E. Forero-Romero]{
  Maria Camila Remolina-Guti\'errez$^{1}$
  \thanks{mc.remolina197@uniandes.edu.co} \&
  Jaime E. Forero-Romero $^{1}$
  \thanks{je.forero@uniandes.edu.co}\\
  $^{1}$ Departamento de F\'isica, Universidad de los Andes, Cra. 1
  No. 18A-10 Edificio Ip, CP 111711, Bogot\'a, Colombia \\
}

\maketitle

\begin{abstract}
Outflows and rotation are two ubiquitous kinematic features in the gas
kinematics of galaxies. 
Here we introduce a semi-analytic model to quantify how rotating
outflows impact the morphology of the Lyman-$\alpha$ emission line.   
The model is contrasted against Monte Carlo radiative transfer
simulations of outflowing gas with additional solid body rotation.
We explore a range of neutral Hydrogen optical depth of
$10^5\leq\tauh\leq10^7$, rotational velocity $0\leq \vrot/\kms \leq
100$ and outflow velocity $0\leq \vout/\kms\leq 50$.  
We find three consequences of rotation.
First, it introduces a dependency with viewing angle; second it
broadens the line and third it increases the flux at the
line's center.
For all viewing angles, the semi-analytic model reproduces the
radiative transfer results for the line width and flux change at the
line's center within a $7\%$ and $50\%$ precision for an optical depth of
$\tauh=10^5$, respectively, and within $2\%$ and $1\%$ for an optical
depth of $\tauh=10^7$.
Using this model we also show that the peaks of integrated spectra
taken from opposite sides of an edge-on rotating gas distribution
should have a separation of $\frac{1}{2}\vrot$. 
The semi-analytic model presented here is a convenient tool to
introduce rotational kinematics as a post-processing step of idealized
Monte Carlo simulations; it provides a framework to interpret \lya
spectra in systems where rotation is expected or directly measured
through kinematic maps.  
\end{abstract}




\section{Introduction}
\label{sec:intro}

Recent advances in instrumentation have revealed the presence of gas
rotation on vastly different physical scales.
For instance, spatially resolved spectra on compact dwarf galaxies
have measured clear signs of gas showing pure rotation kinematics
\citep{2015A&A...577A..21C,2017A&A...600A.125C} and the recent mapping of
high redshift circumgalactic regions have also revealed kinematic
evidence for large scale rotation \citep{2018MNRAS.473.3907A}.
Systems with star formation, neutral gas and low dust
contents can produce a \lya emission line \citep{PartridgePeebles}
motivating the observational work to phenomenologically link tracers
of galaxy rotation such as H$\alpha$ to \lya spectra
\citep[e.g.][]{Herenz2016}.  

What is then the expected imprint of rotation on a resonant emission
line such as the \lya line? To what extent is it possible to constrain
rotational kinematics from the \lya emission line? 
Detailed radiative transfer (RT) \lya modeling of rotating systems
started until recently by \cite{Garavito14}.
In that work the authors studied the influence of pure solid body
rotation on the \lya line's morphology.
They found that rotation indeed introduces changes, the most
noticeable being the dependence of the spectra with the viewing
angle with respect to the rotation axis. 

\cite{Garavito14} also presented a simple semi-analytical
approximation that accounted for the main features of the \lya spectra
from a rotation sphere.
Recently, this semi-analytic solution was used to perform a Markov
Chain Monte Carlo exploration to fit the observed spectra Compact
Dwarf Galaxy \citep{tololo} with atypical features that could be
explained by pure rotation.  

However, the gas dynamics in Lyman Alpha Emitter (LAE) galaxies are
more complex than pure rotation.
In many observations the \lya line profile has a single peak
redwards from the line's center, in other cases there is a double peak
but the peak on the red side is stronger
\citep[e.g.][]{2010ApJ...717..289S,Erb14,Trainor16}.   
These features have been explained as the consequence of multiple
\lya photon scatterings through a homogeneous outflowing shell of
neutral Hydrogen
\citep{Verhamme06,Orsi12,2015ApJ...812..123G}.  

Nevertheless, a study of the combined effects of outflows and
rotation has not been presented in the literature.
Here we report on such a study with the main aim of quantifying the
validity of the semi-analytic approximation presented by
\cite{Garavito14} in the case where outflows are also present. 
We investigate a simplified geometrical configuration corresponding to a
spherical gas cloud with symmetrical radial outflows and solid body
rotation and contrast the semi-analytic model against the results of a
Monte-Carlo radiative transfer code. 

The structure of the paper is the following.
We introduce first our theoretical tools and assumptions
in Section \ref{sec:theory}. We continue in Section \ref{sec:results}
with the results from the Monte-Carlo simulation, the comparison
against the semi-analytical approximation which we use to make a
thorough exploration of the effect of rotation.
In Section \ref{sec:discussion} we discuss our results and their
possible implications for observational analysis to finally present
our conclusions in Section \ref{sec:conclusions}.

Throughout the paper we use a thermal velocity for a neutral Hydrogen
gas of \vth $= 12.86$ \kms, which corresponds to a temperature of
$T=10^4$ K. 

\begin{figure*}
\centering
    \includegraphics[width=0.48\textwidth]{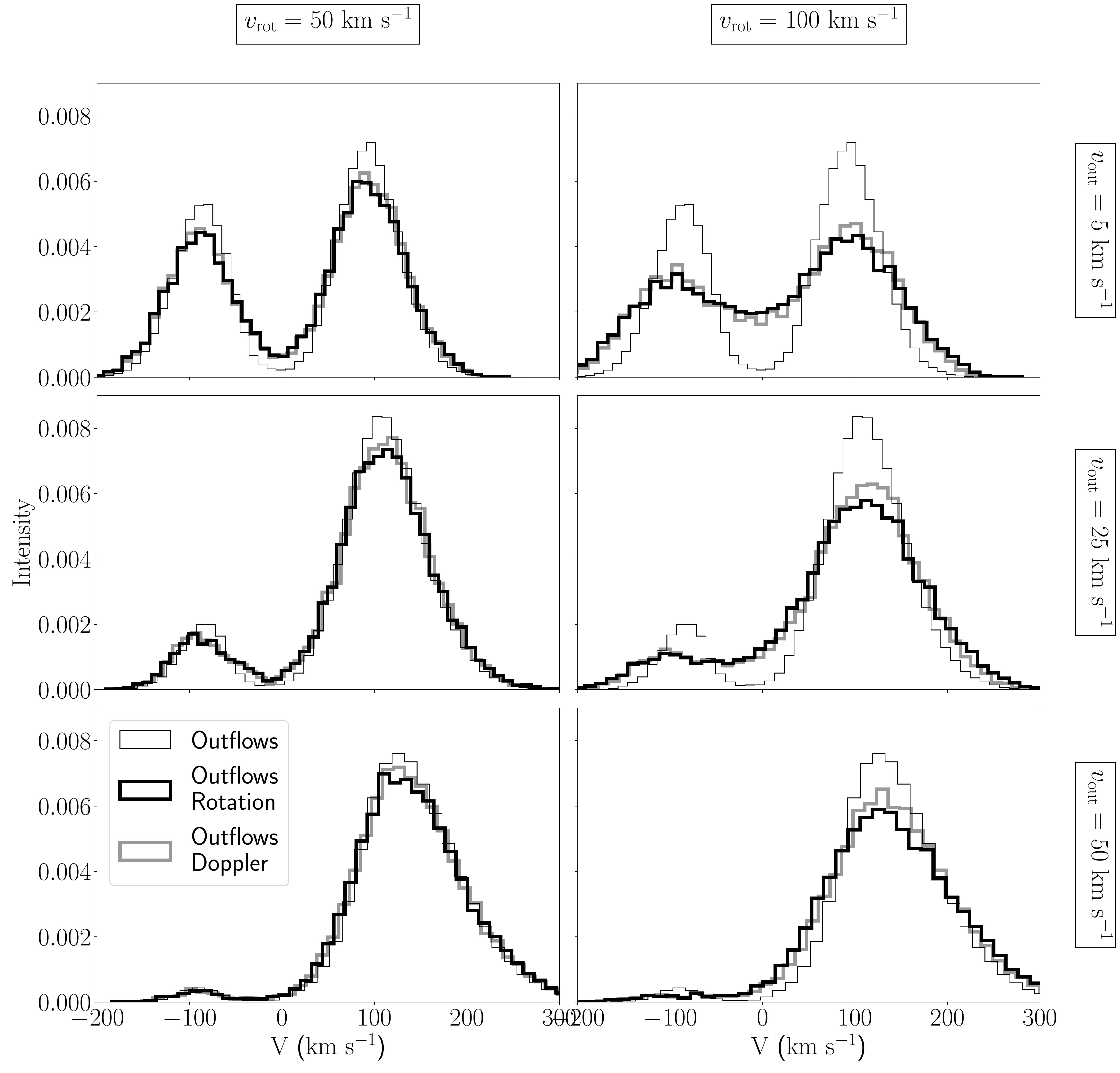}
    \includegraphics[width=0.48\textwidth]{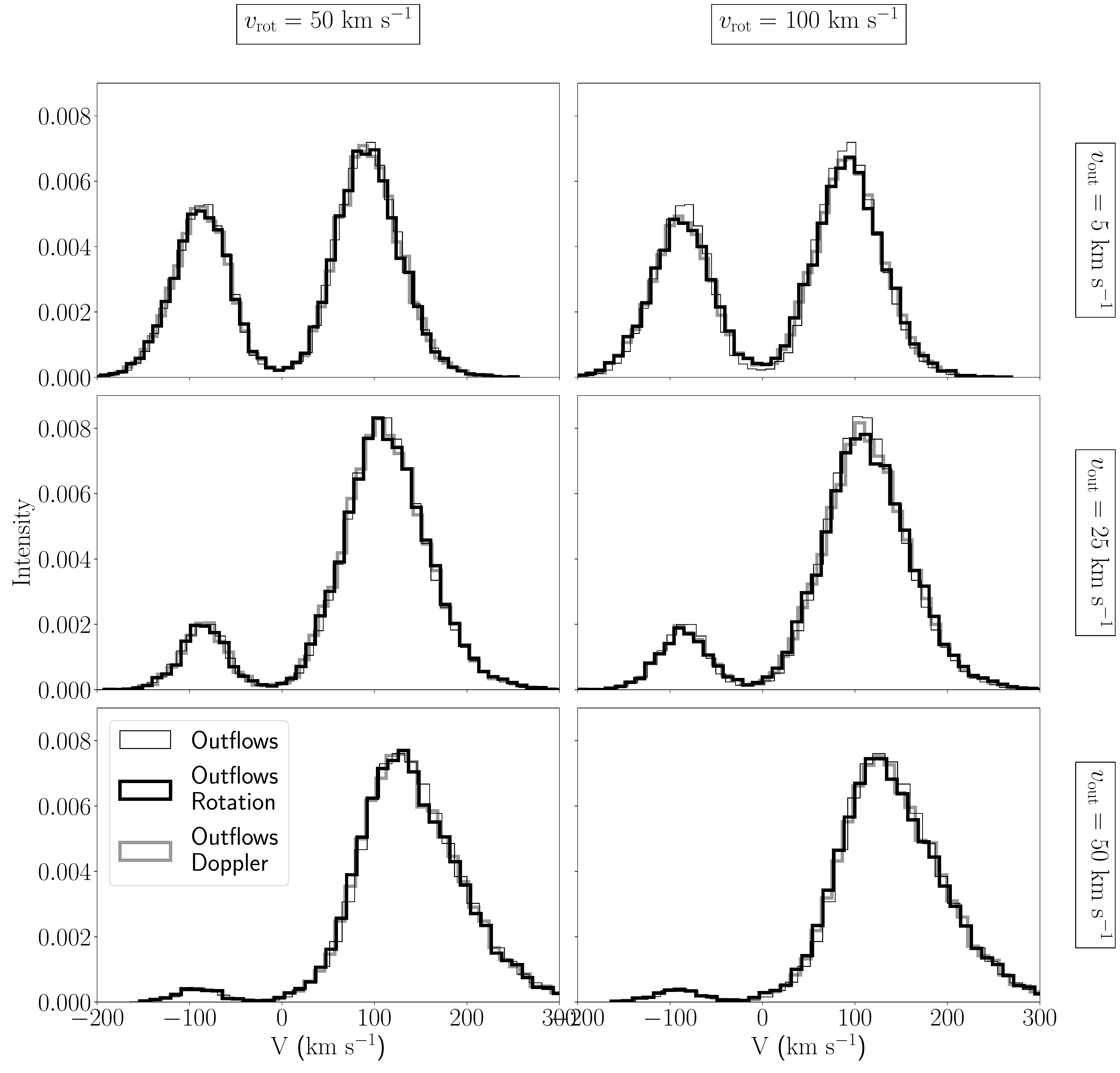}
  \caption{\textbf{Qualitative trends of changing outflow and
      rotational velocity viewed perpendicular/paralell to the
      rotation axis}.  
    Here we fix $\tauh=10^6$. 
    The six panels on the left correspond to $\theta=90^\circ$ and the
    panels on the right to $\theta=0^{\circ}$
    We vary \vrot increasing from left to right and \vout increasing
    from top to bottom. 
    The thin black line corresponds to the \lya line obtained with
    CLARA without any rotation and the indicated outflow velocity.
    The thick black line corresponds to CLARA's results including both
    outflows and rotation.
    The thick gray line shows the results of modifying the pure outflow
    solution (thin black line) by the Doppler shift presented in Equation \ref{eq:shift_x} using the respective \vrot. 
    \label{fig:doppler_shift}}
\end{figure*}

\begin{figure*}
\begin{center}
\includegraphics[height=0.25\textwidth]{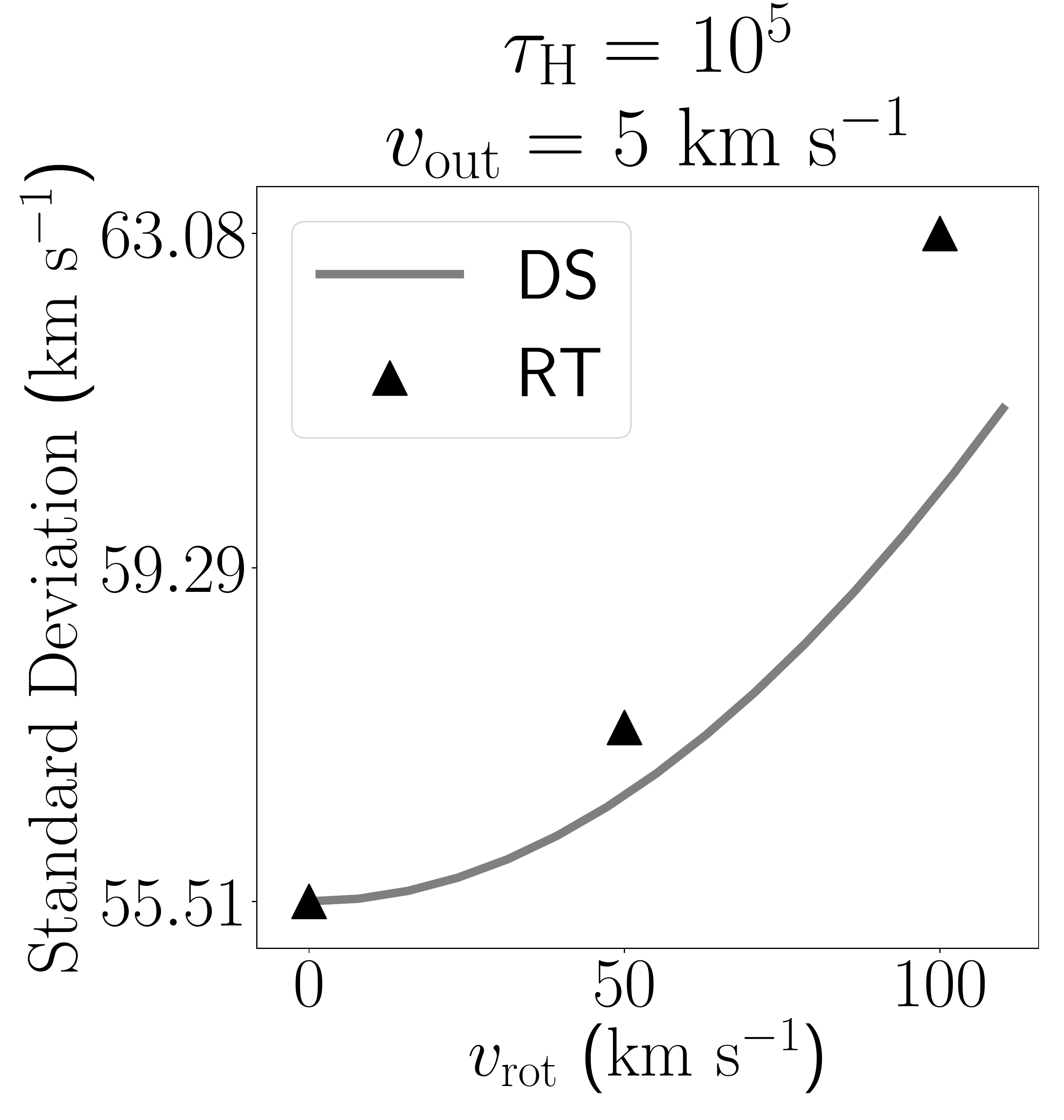}
\includegraphics[height=0.25\textwidth]{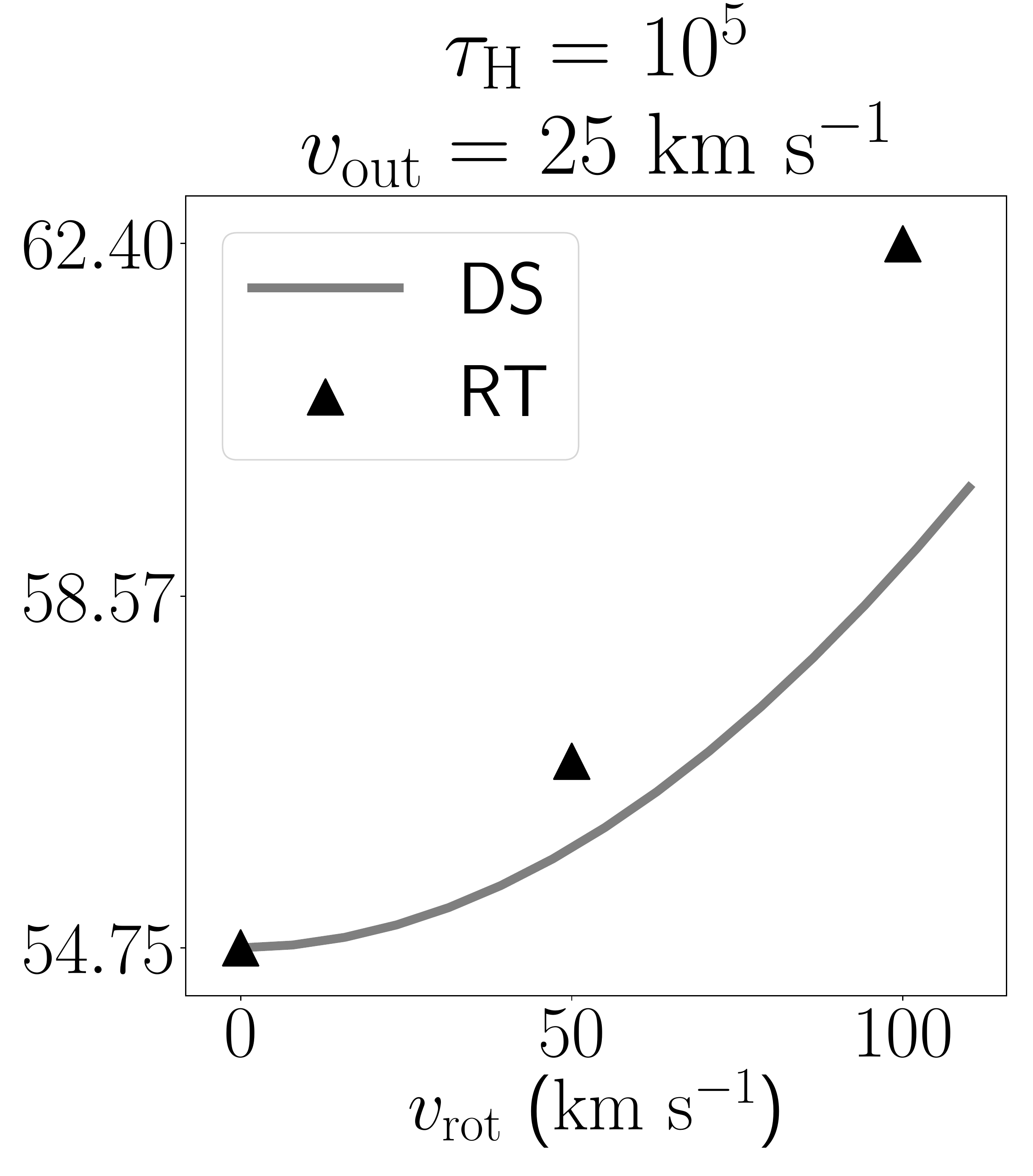}
\includegraphics[height=0.25\textwidth]{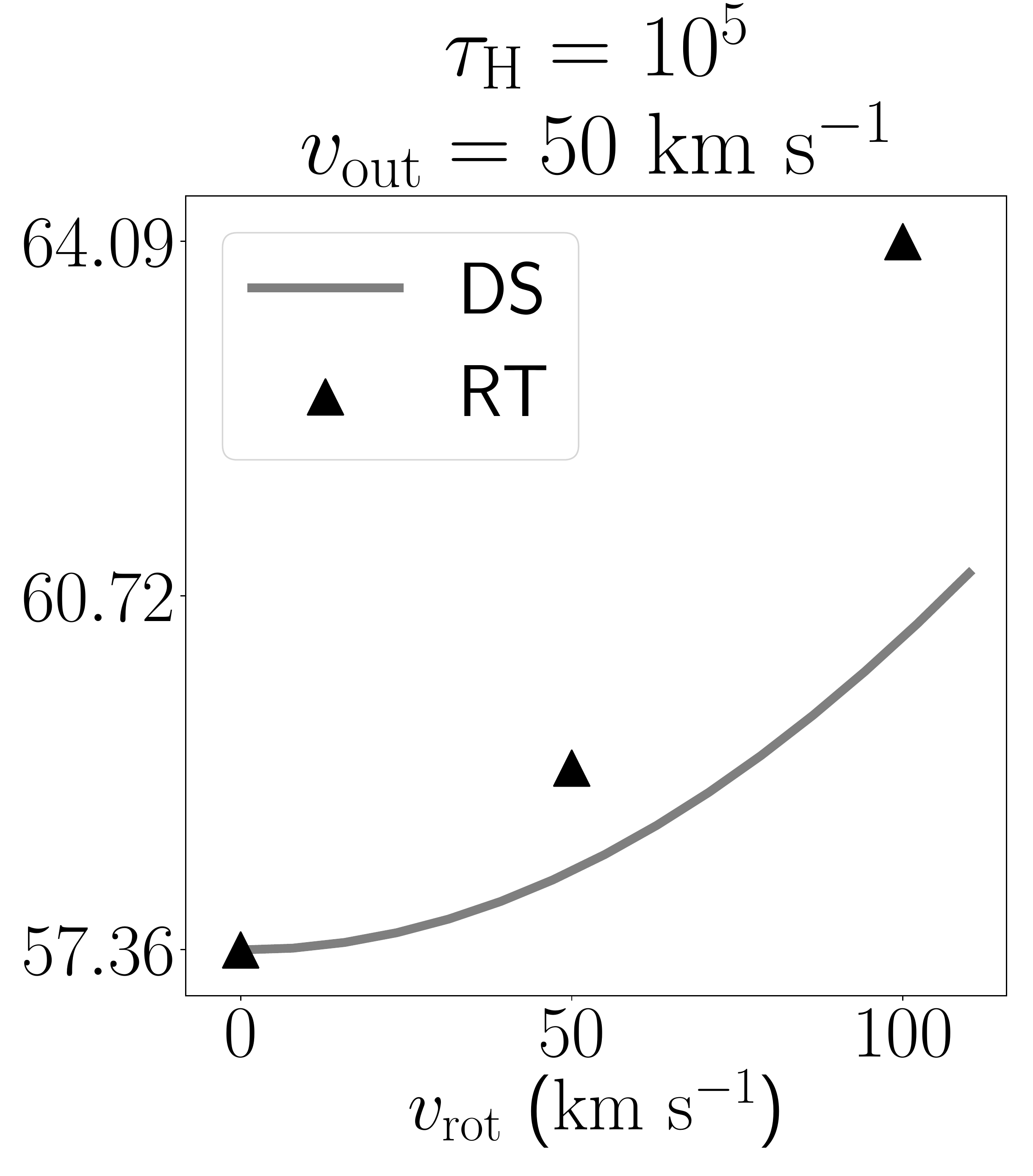}\\
\includegraphics[height=0.25\textwidth]{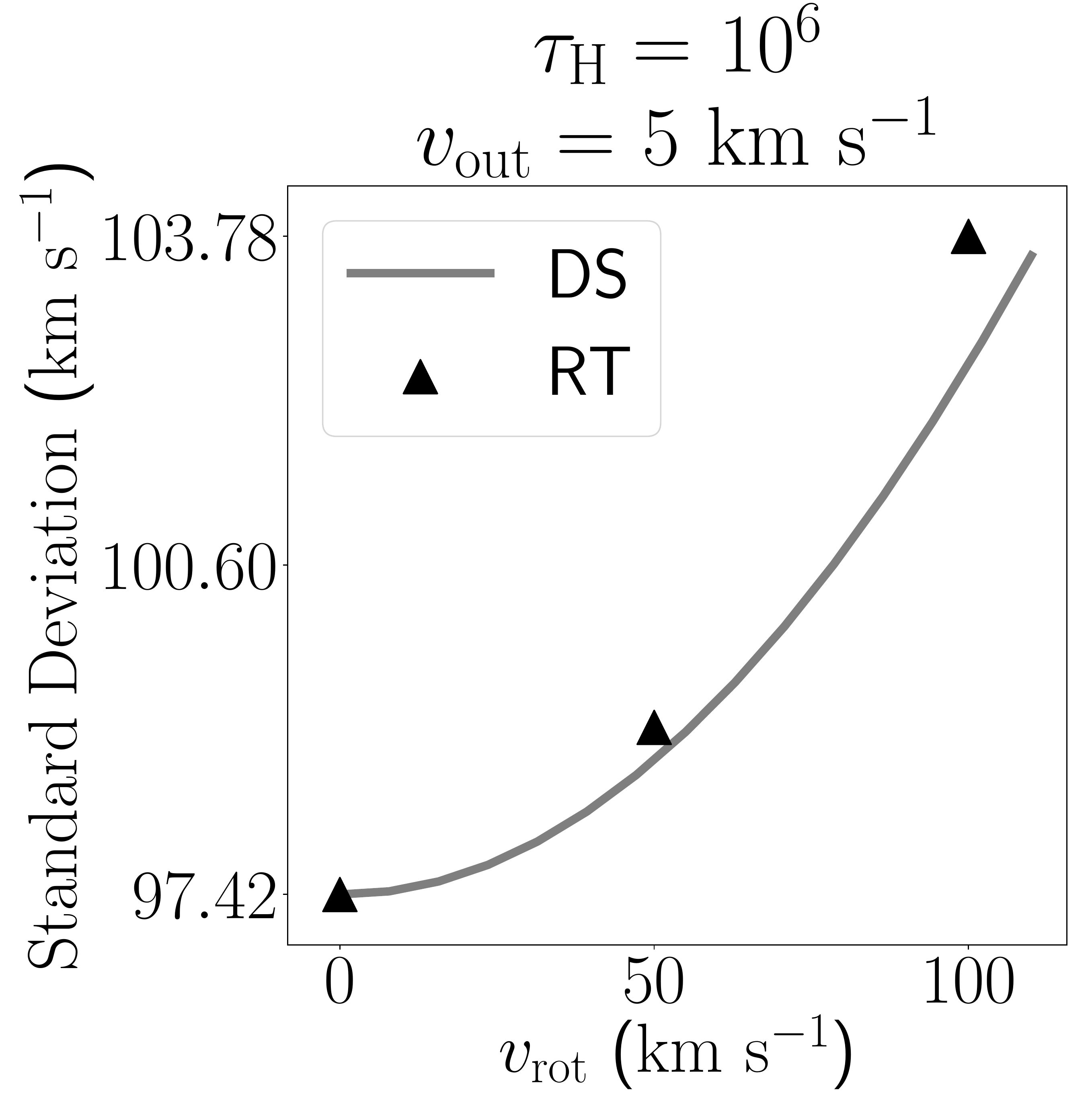}
\includegraphics[height=0.25\textwidth]{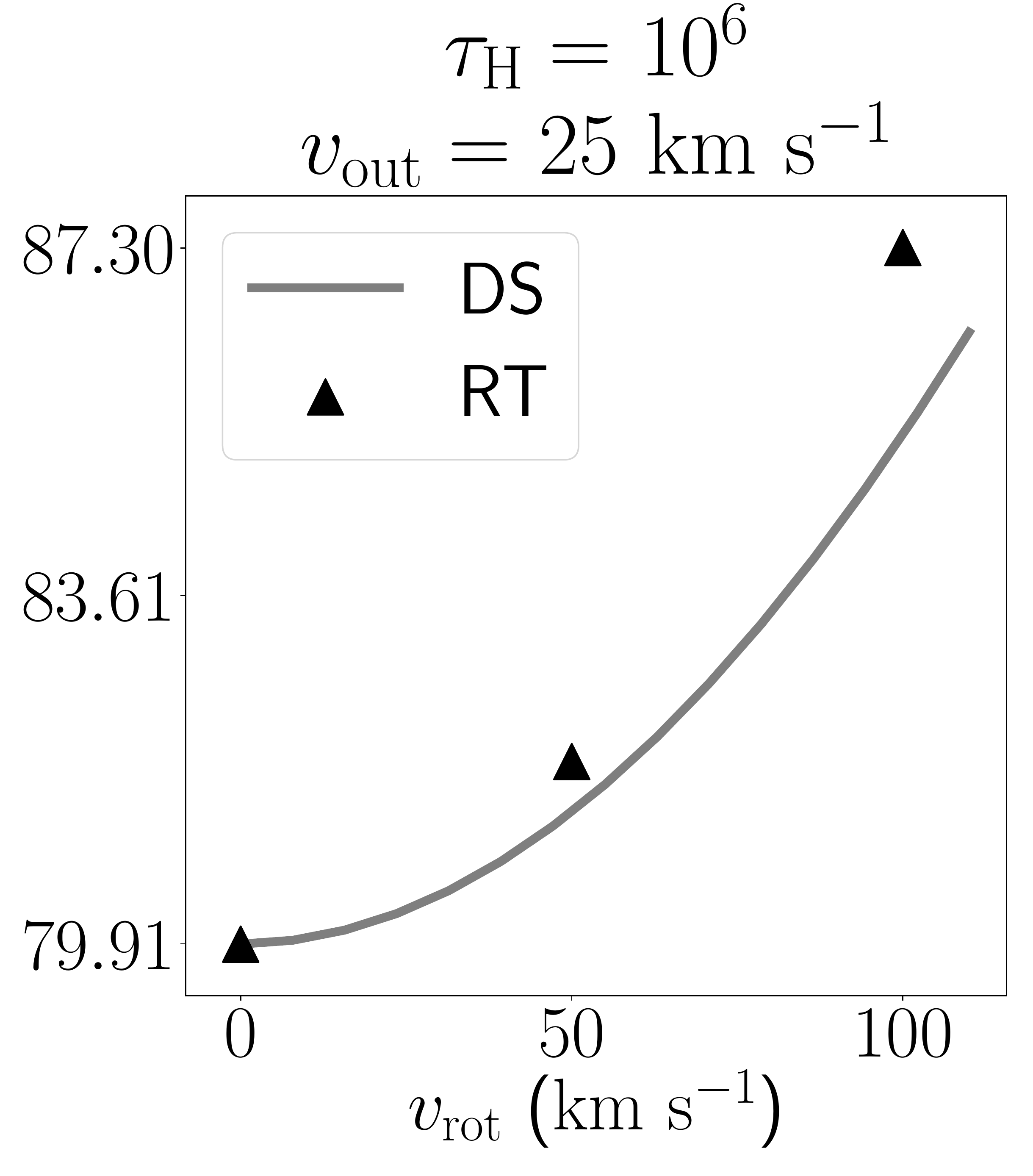}
\includegraphics[height=0.25\textwidth]{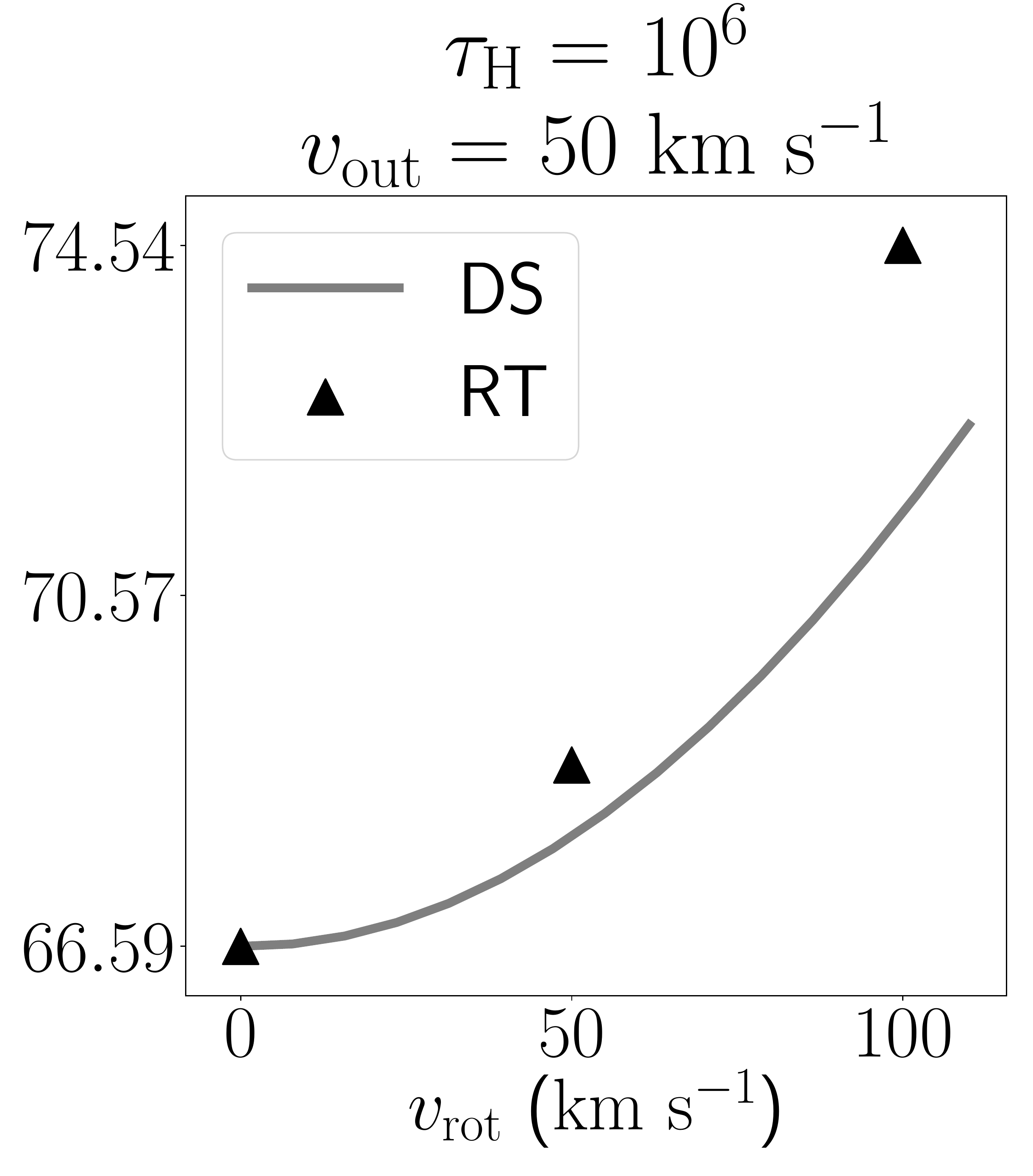}\\
\includegraphics[height=0.25\textwidth]{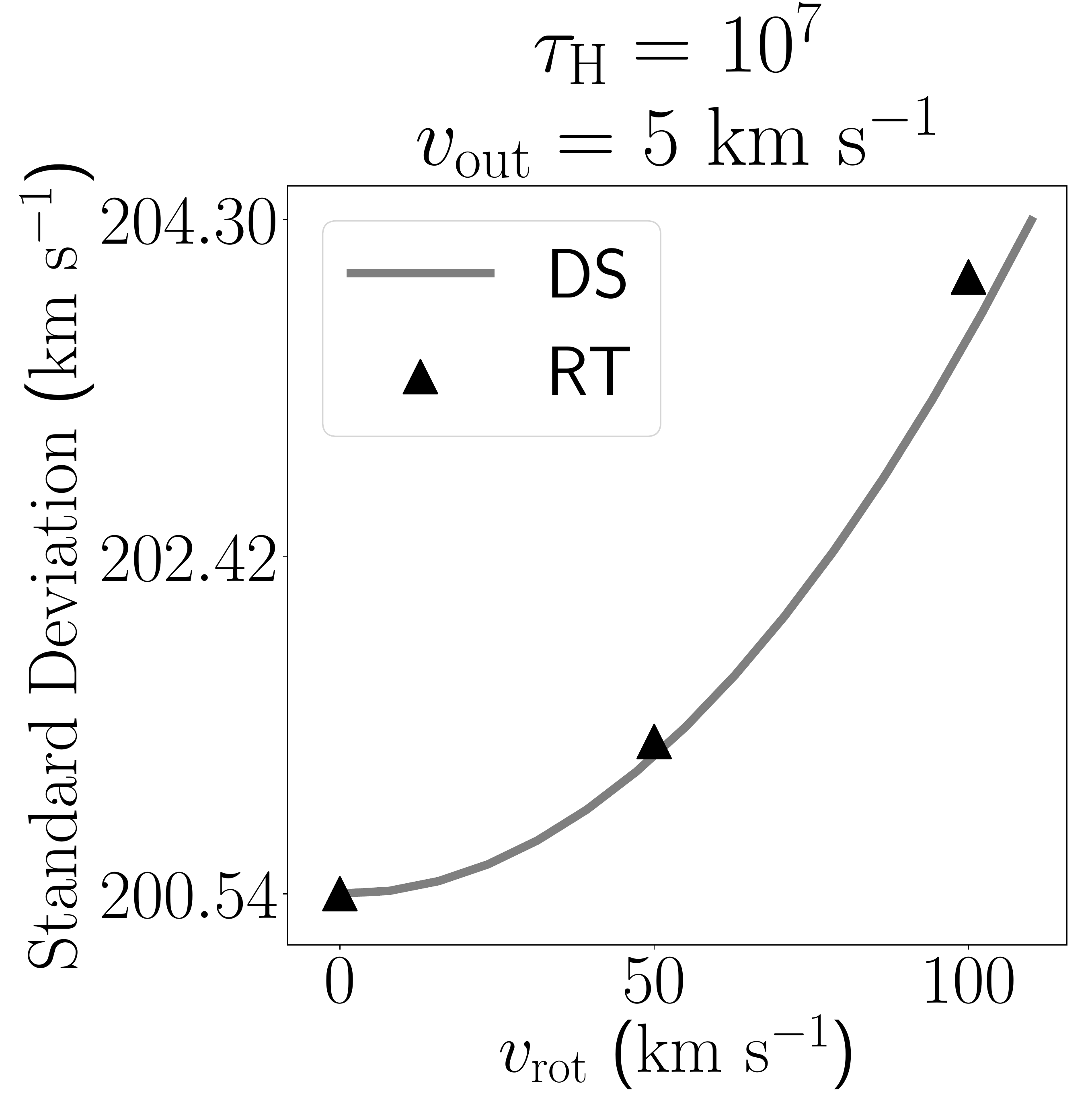}
\includegraphics[height=0.25\textwidth]{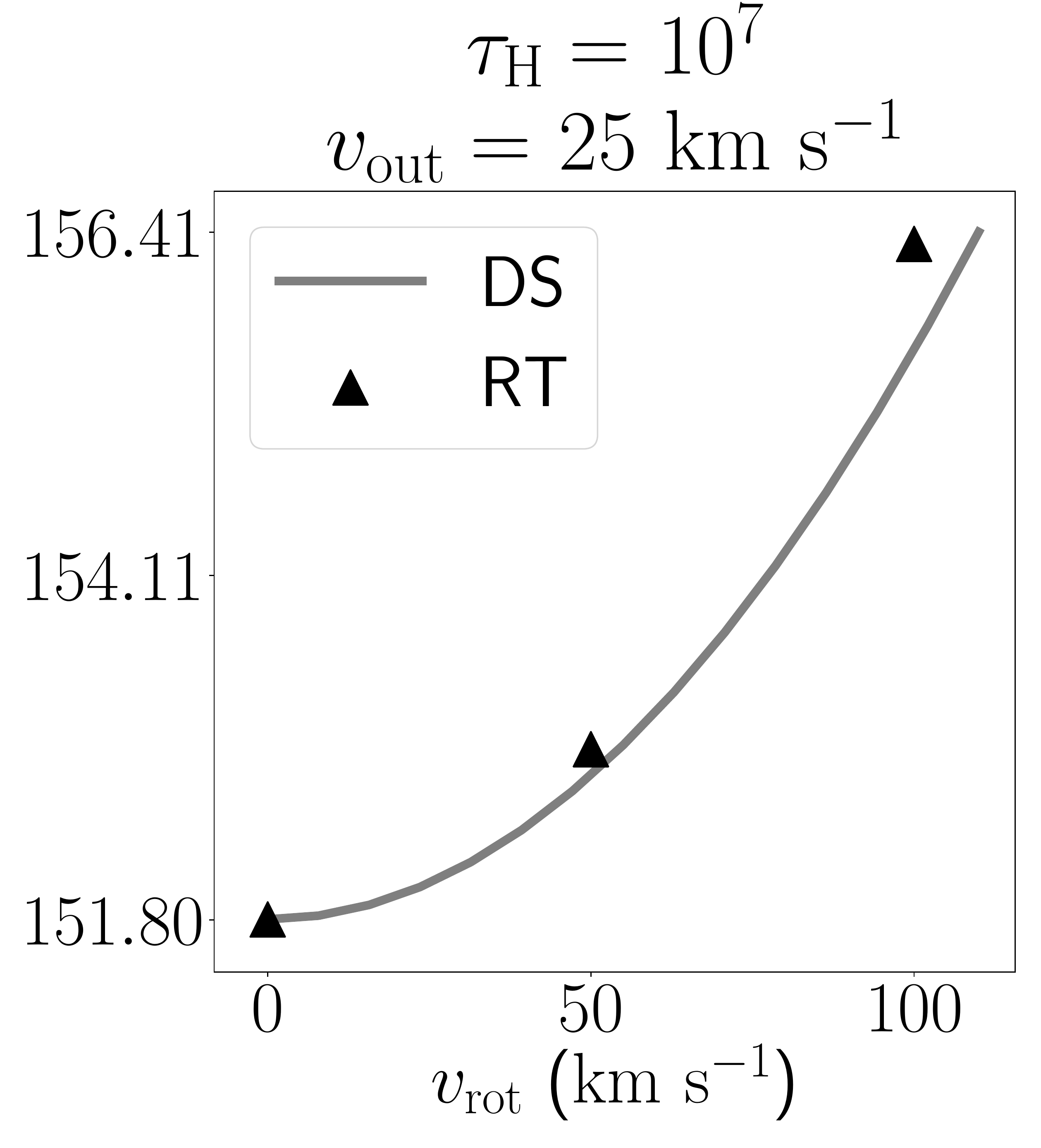}
\includegraphics[height=0.25\textwidth]{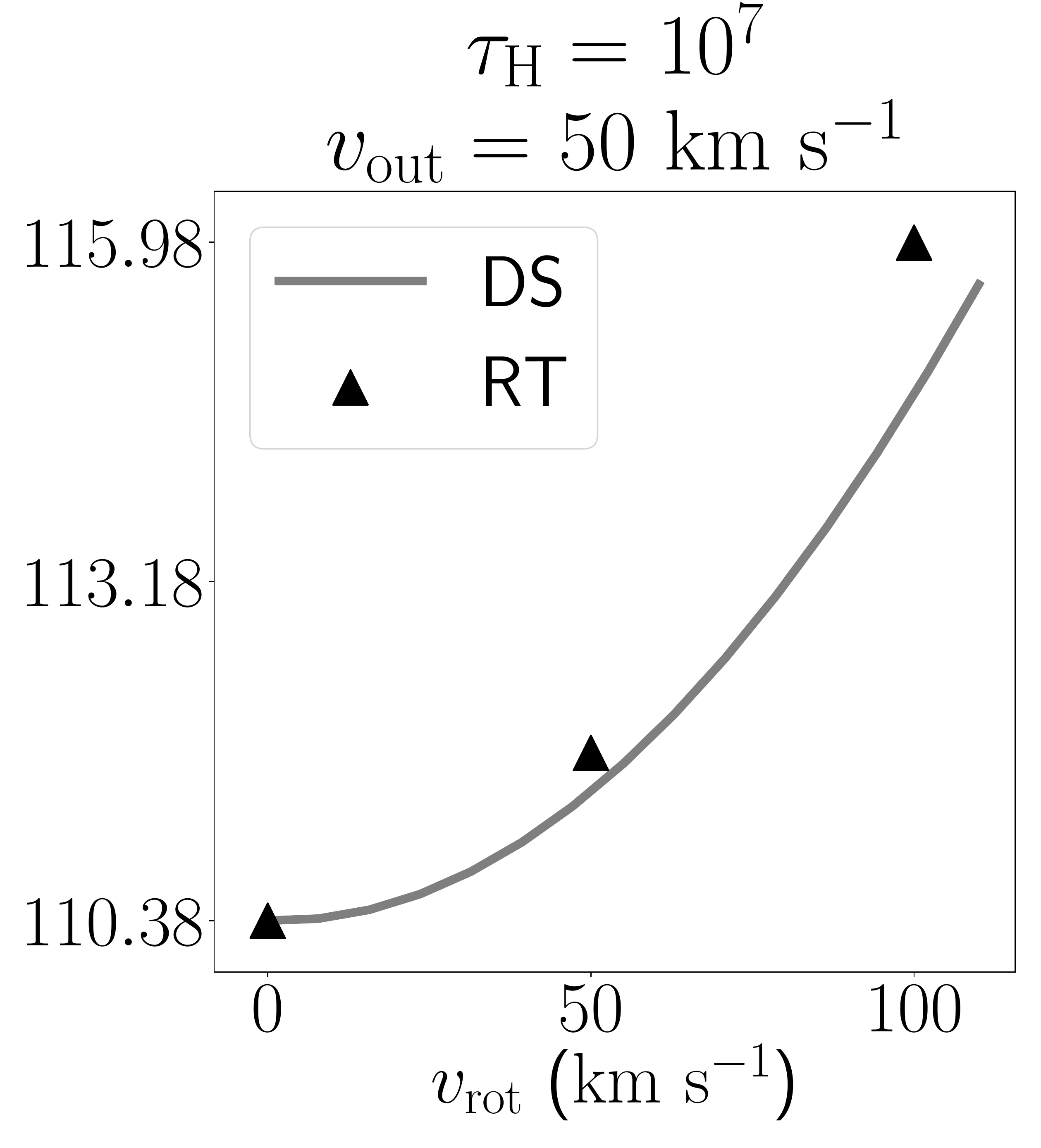}
\end{center}
\caption{\textbf{Standard Deviation trends.} Results for all the
  Radiative Transfer simulations (in triangles) compares against the
  Doppler Shift model (lines).
  All panels correspond to a viewing angle of $\theta = 90^{\circ}$
  (perpendicular to the rotation axis). 
  The optical depth increases from top to bottom and the outflow
  velocity from left to right.
  \label{fig:standard_deviation}}
\end{figure*}

\begin{figure*}
\begin{center}
\includegraphics[height=0.25\textwidth]{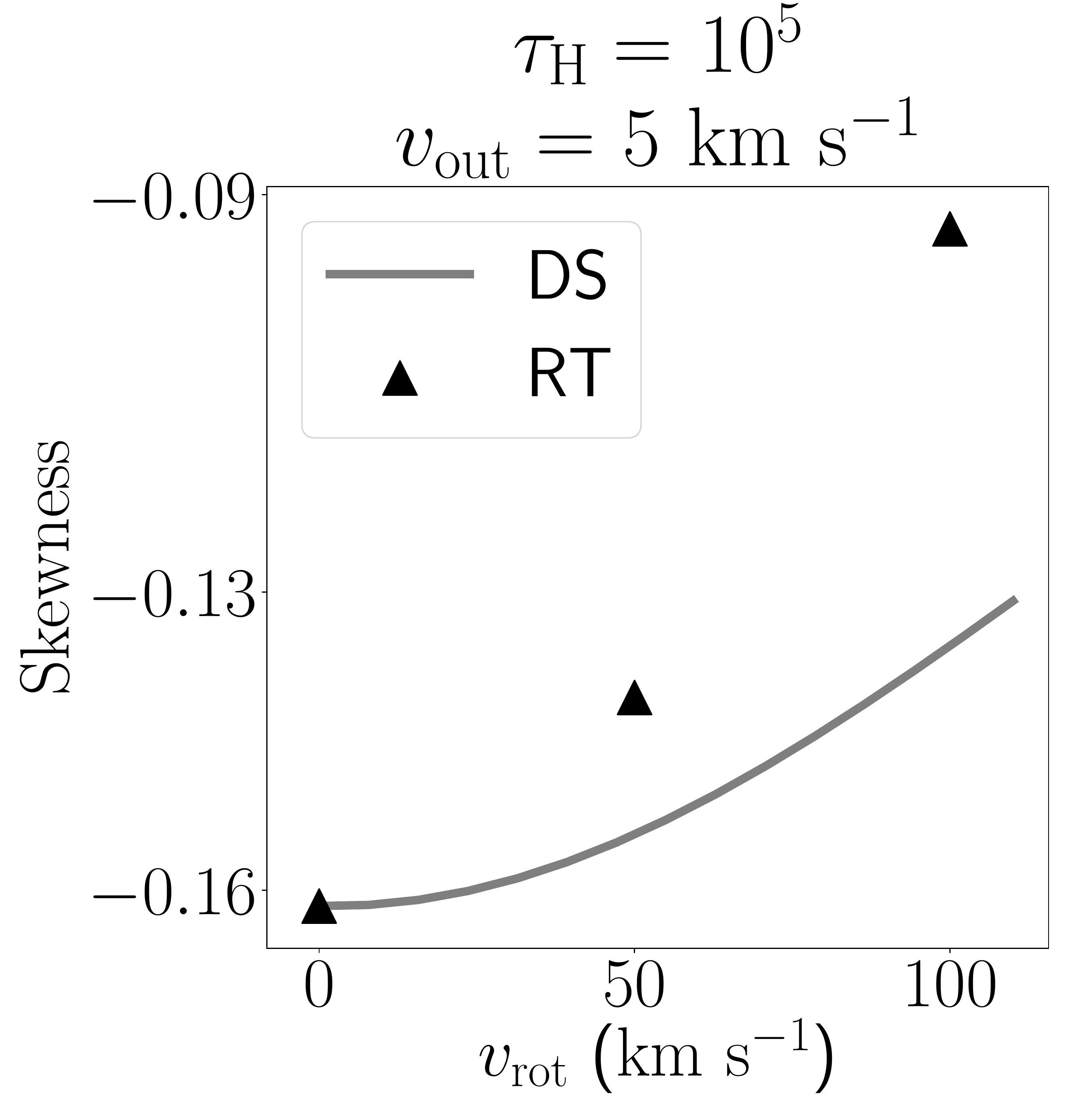}
\includegraphics[height=0.25\textwidth]{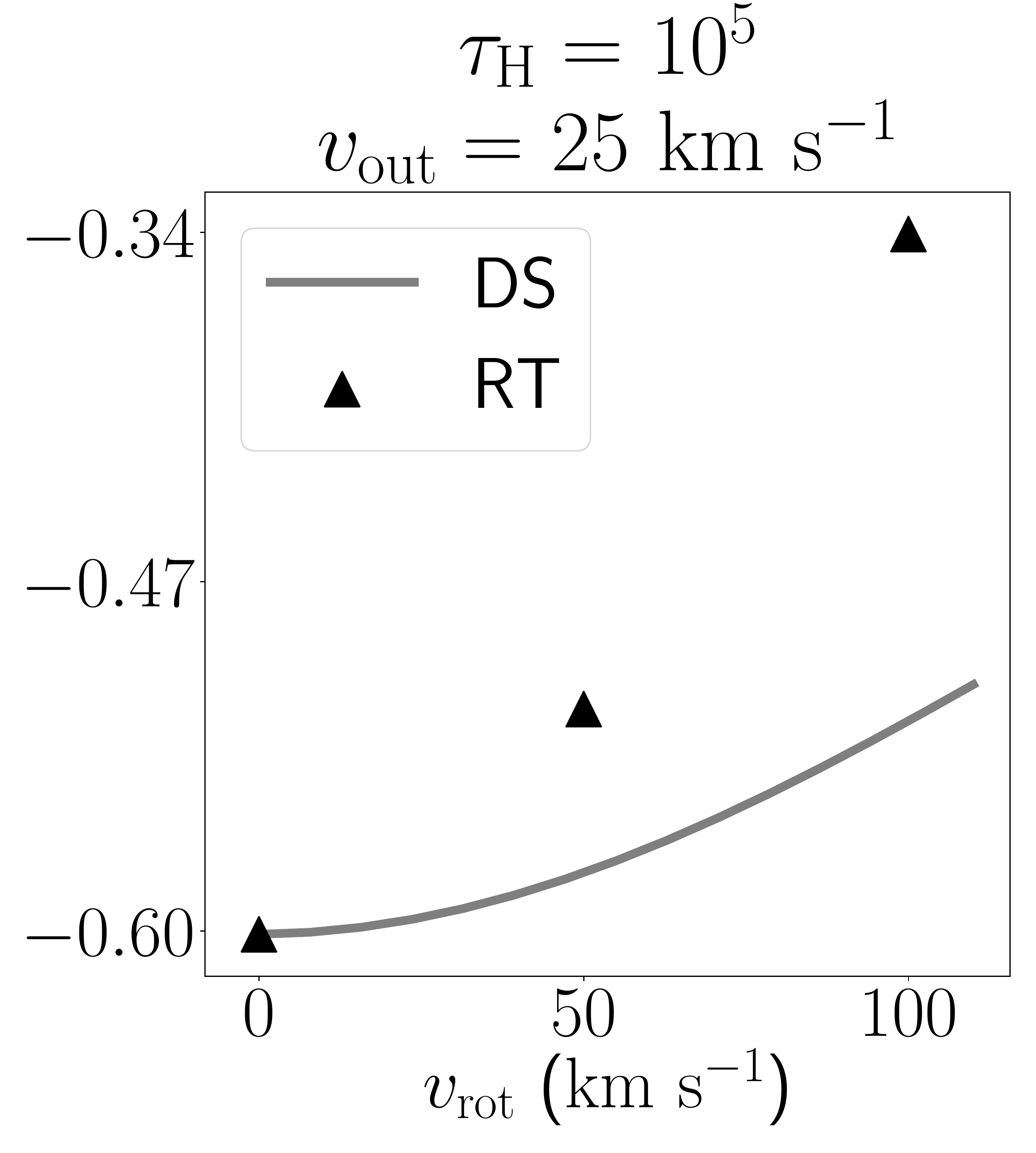}
\includegraphics[height=0.25\textwidth]{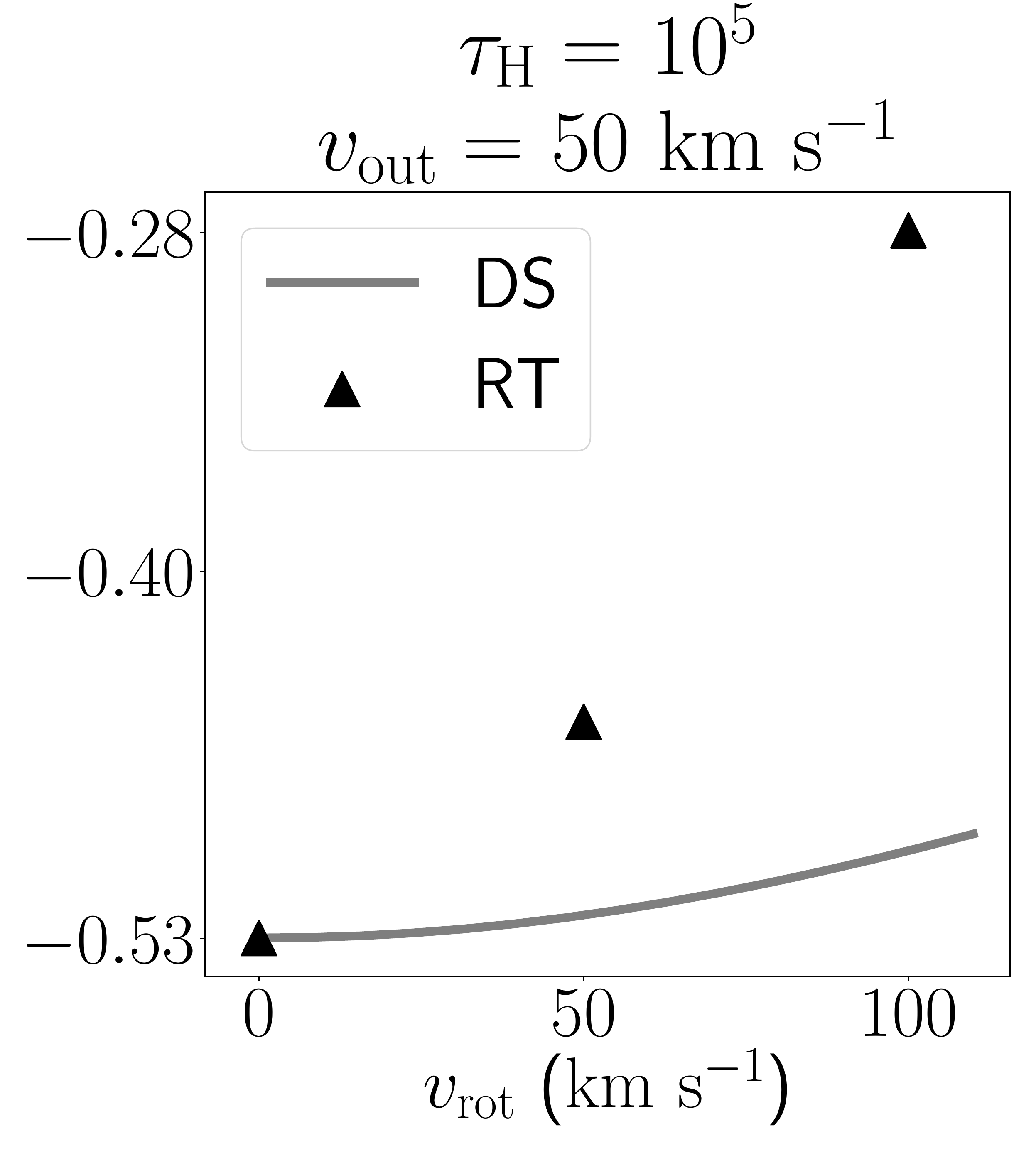}\\
\includegraphics[height=0.25\textwidth]{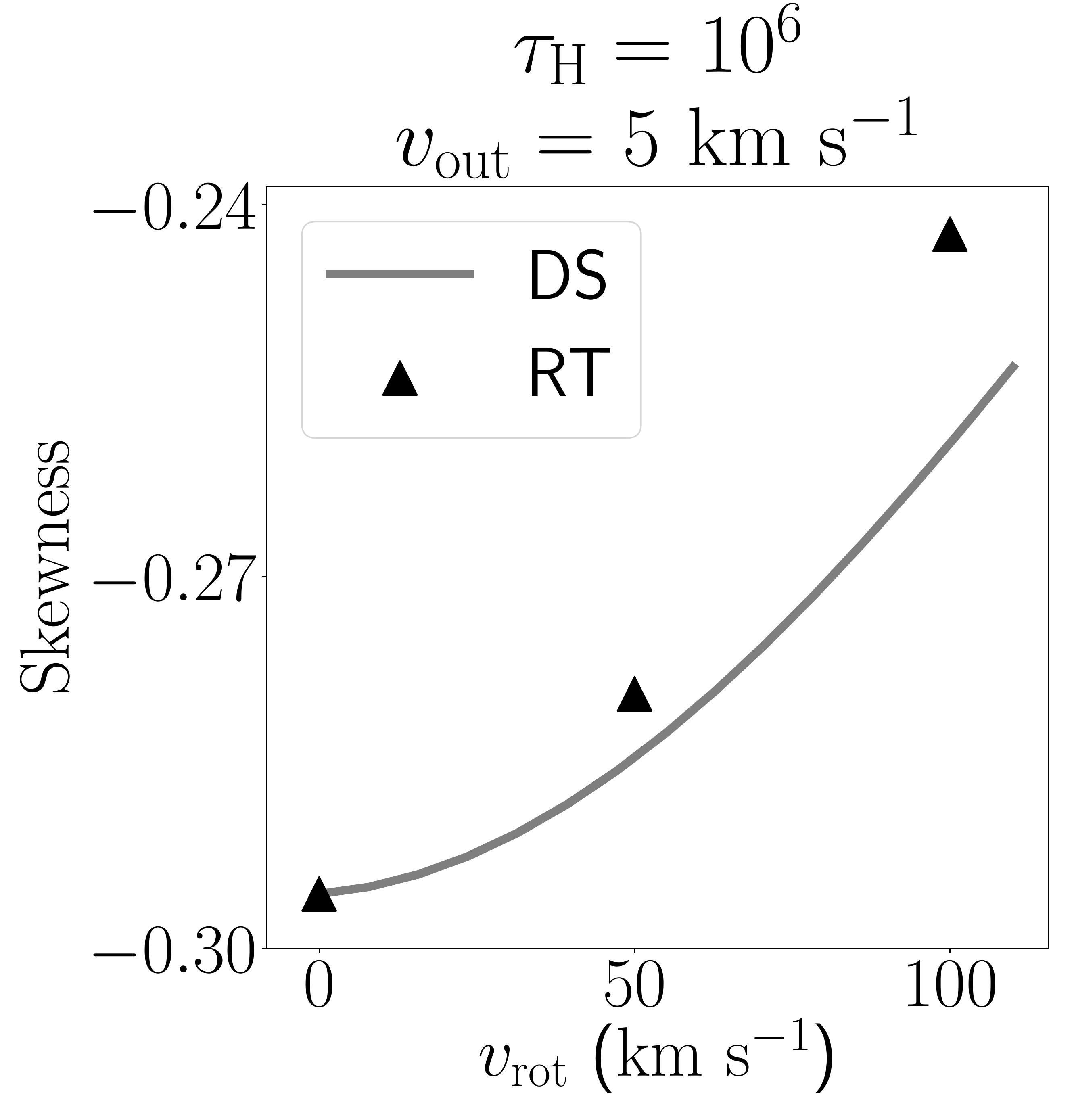}
\includegraphics[height=0.25\textwidth]{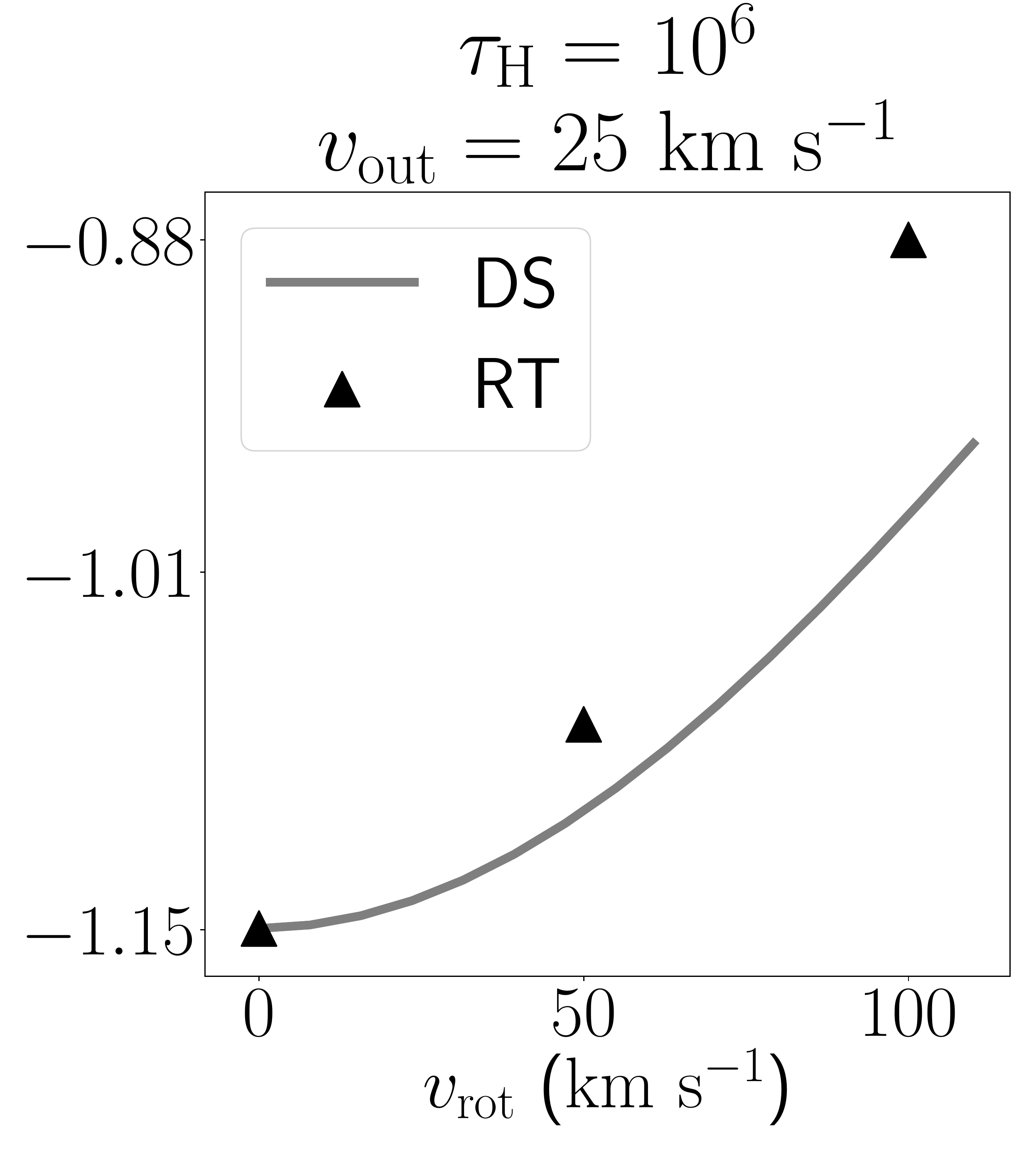}
\includegraphics[height=0.25\textwidth]{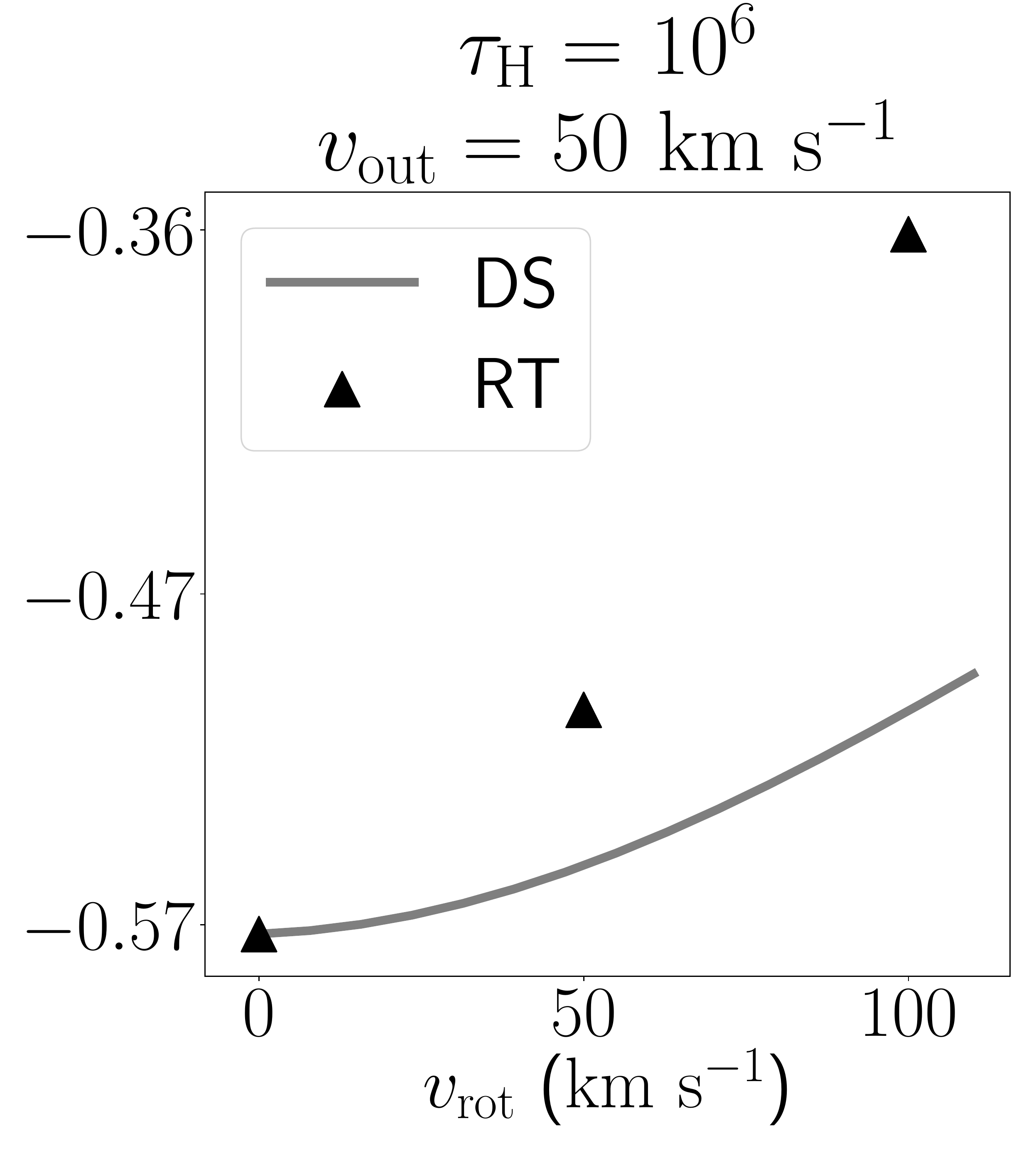}\\
\includegraphics[height=0.25\textwidth]{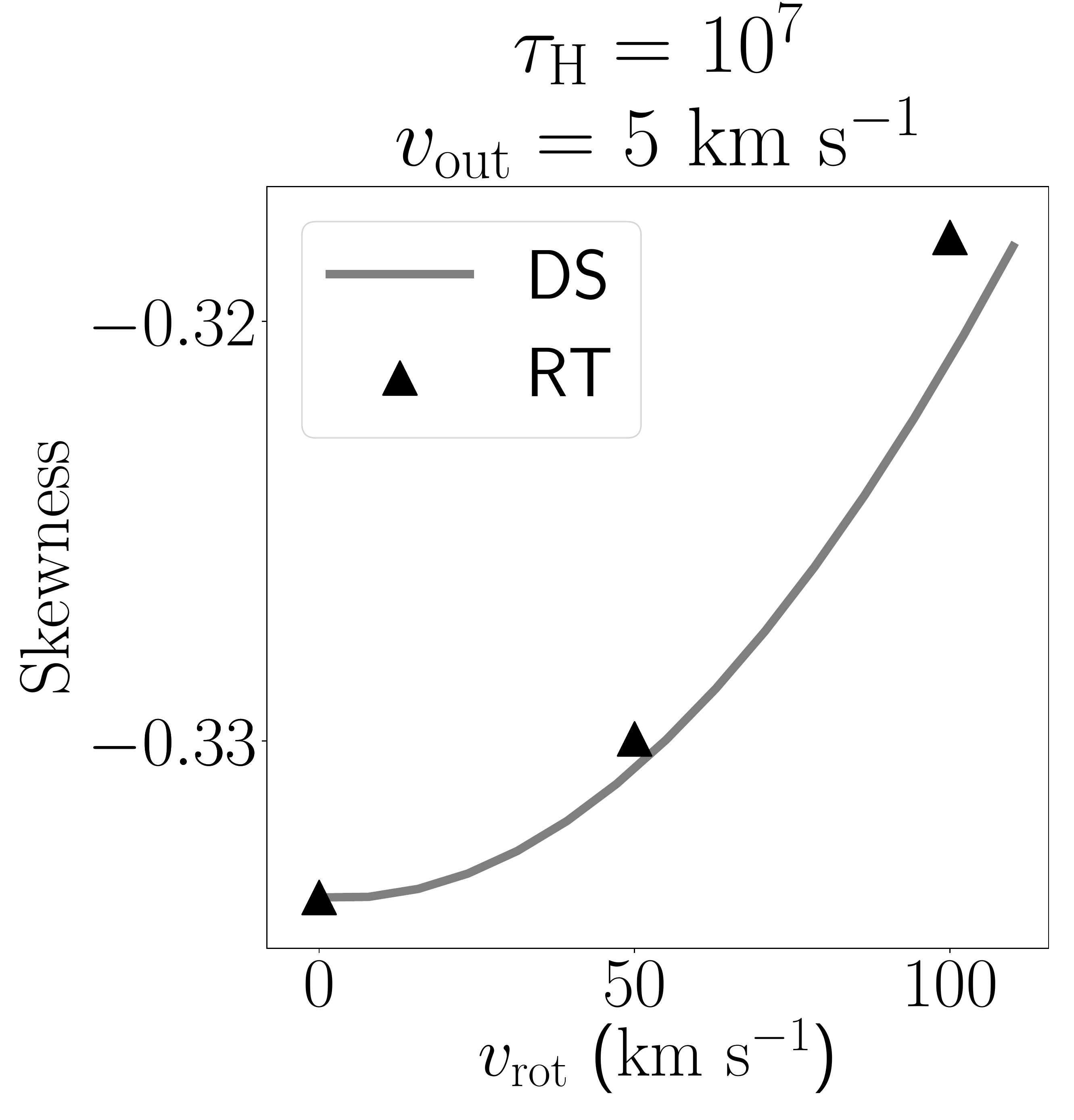}
\includegraphics[height=0.25\textwidth]{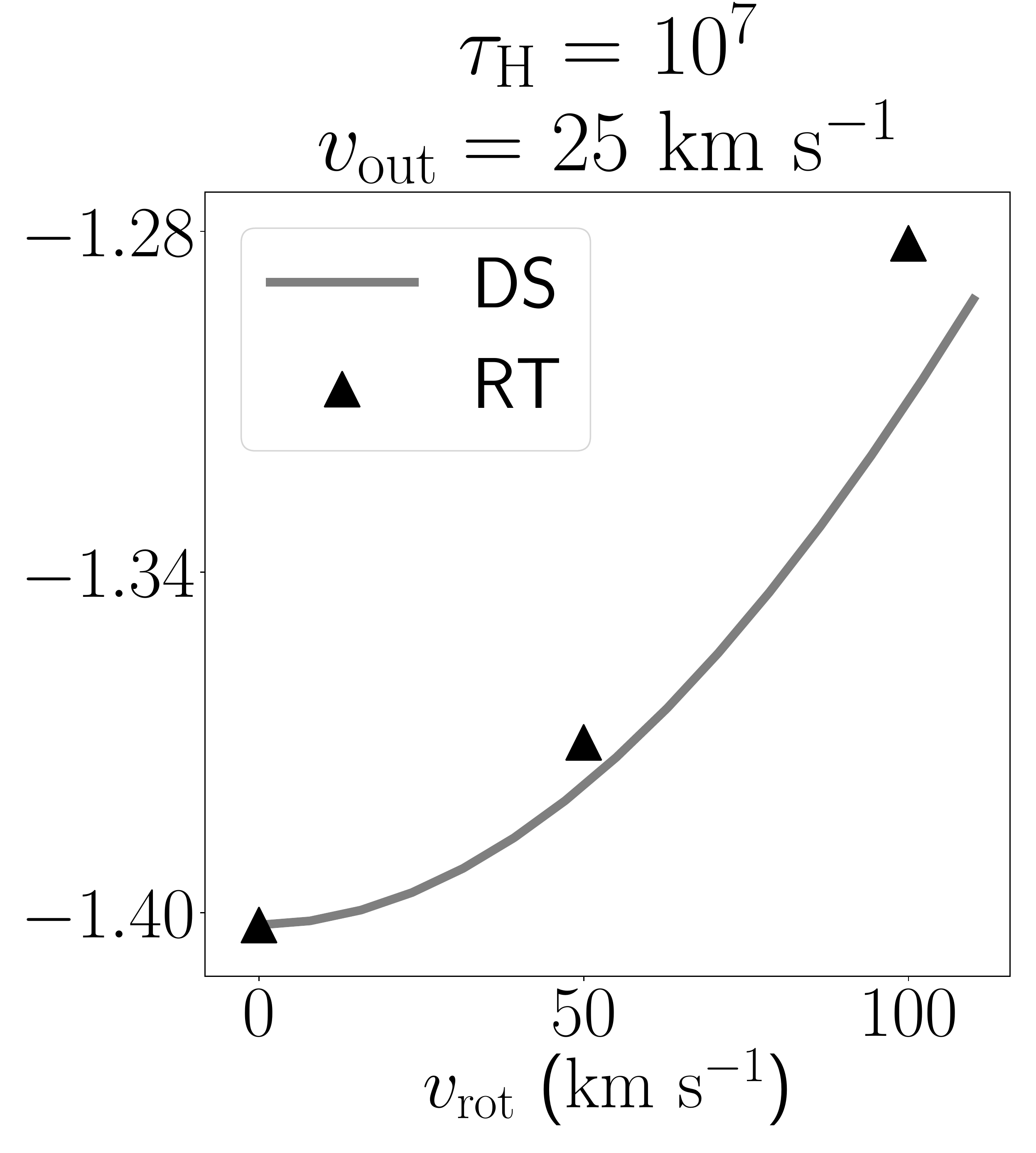}
\includegraphics[height=0.25\textwidth]{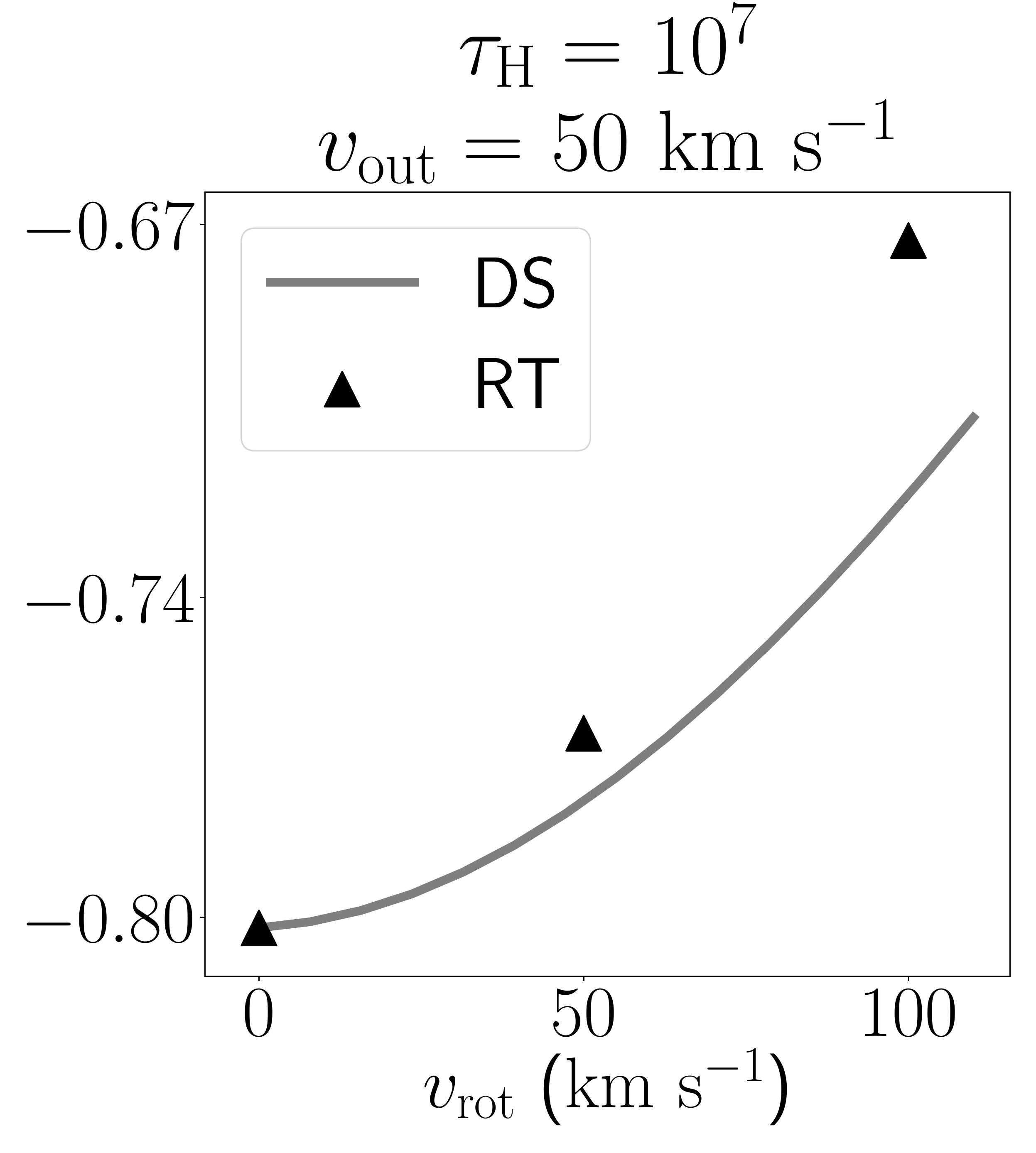}
\end{center}
\caption{\textbf{Skewness trends.} Results for all the
  Radiative Transfer simulations (in triangles) compares against the
  Doppler Shift model (lines).
  Follows the same layout as Figure \ref{fig:standard_deviation}. 
  \label{fig:skewness}}
\end{figure*}

\begin{figure*}
\begin{center}
\includegraphics[height=0.25\textwidth]{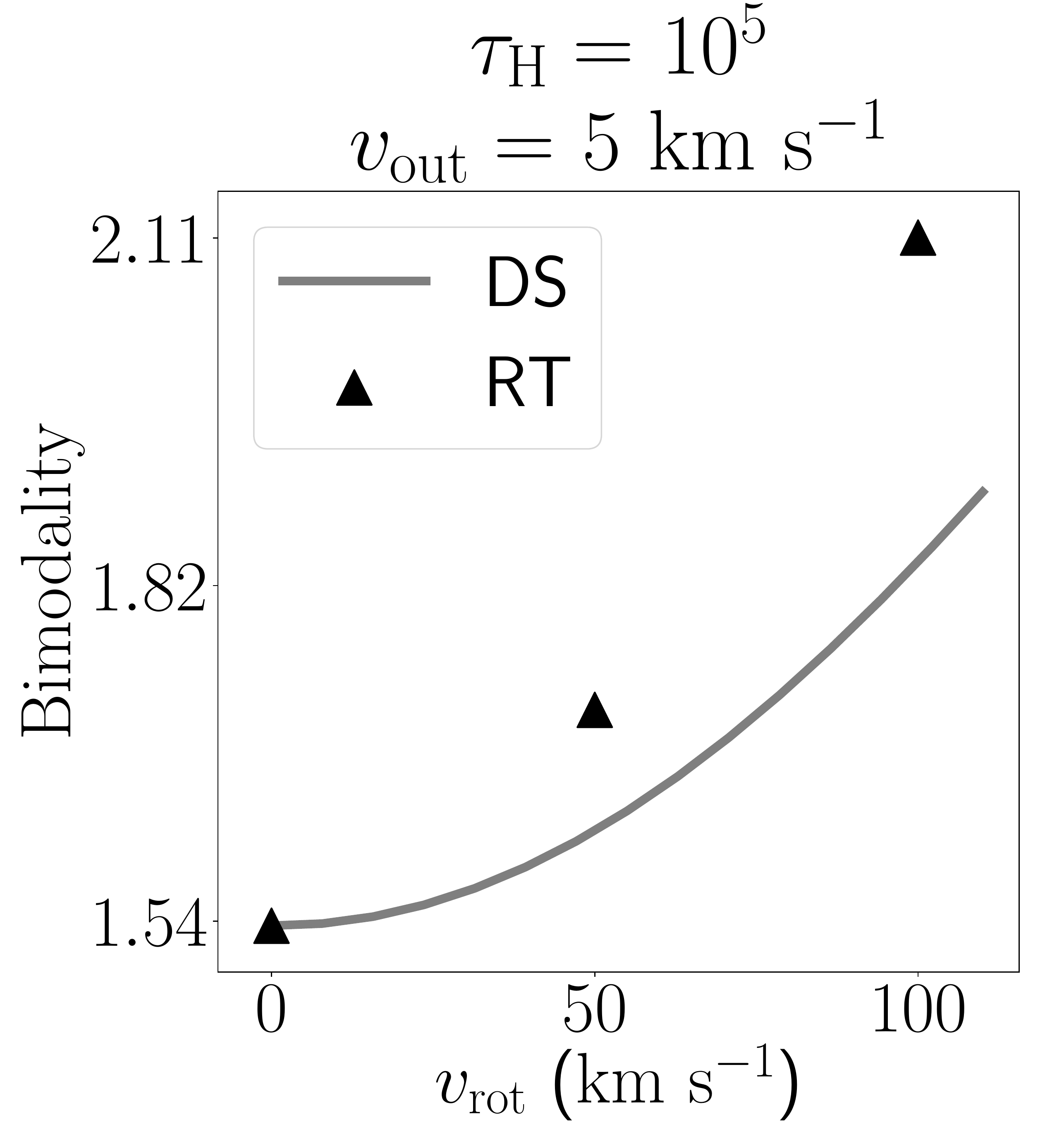}
\includegraphics[height=0.25\textwidth]{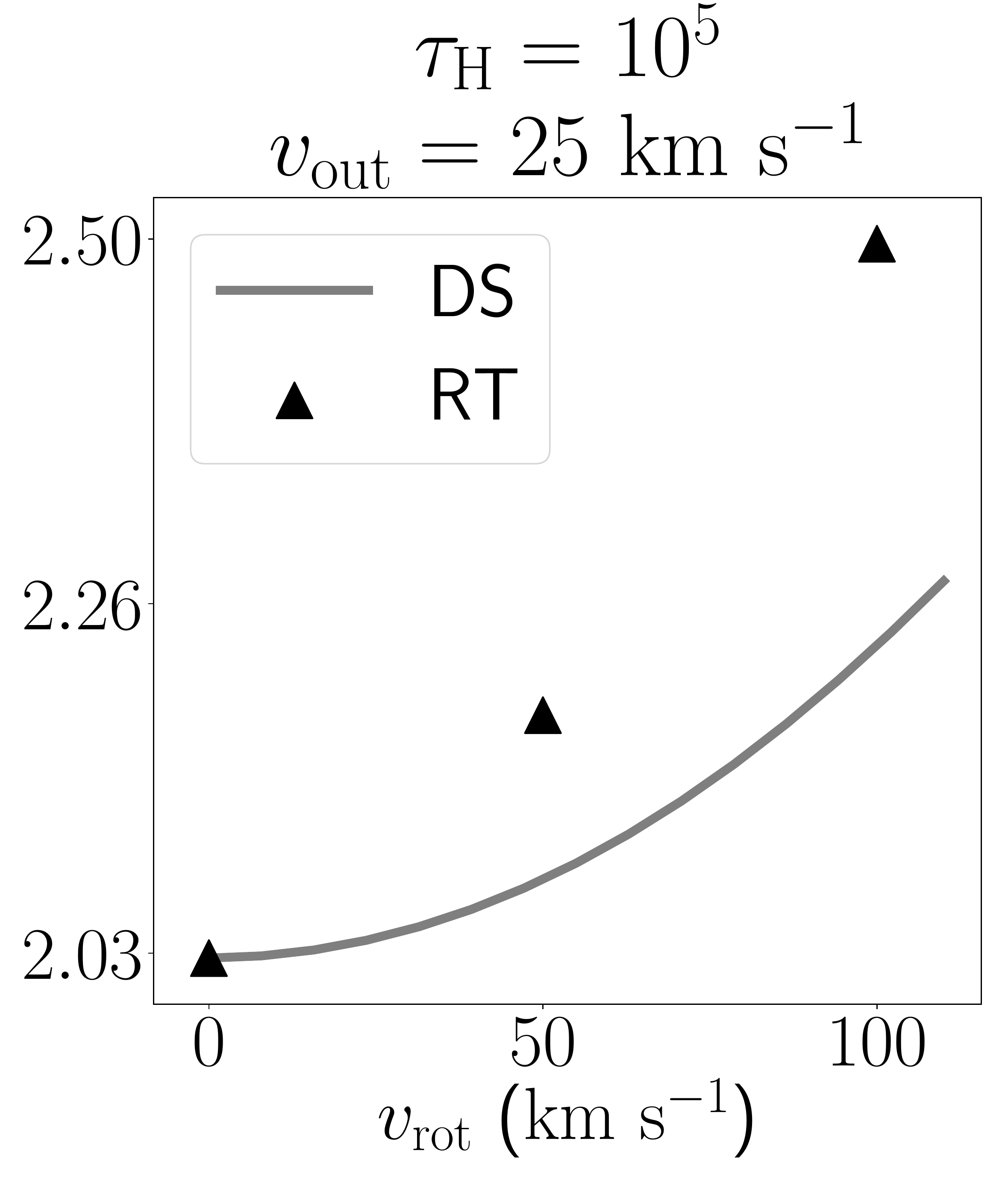}
\includegraphics[height=0.25\textwidth]{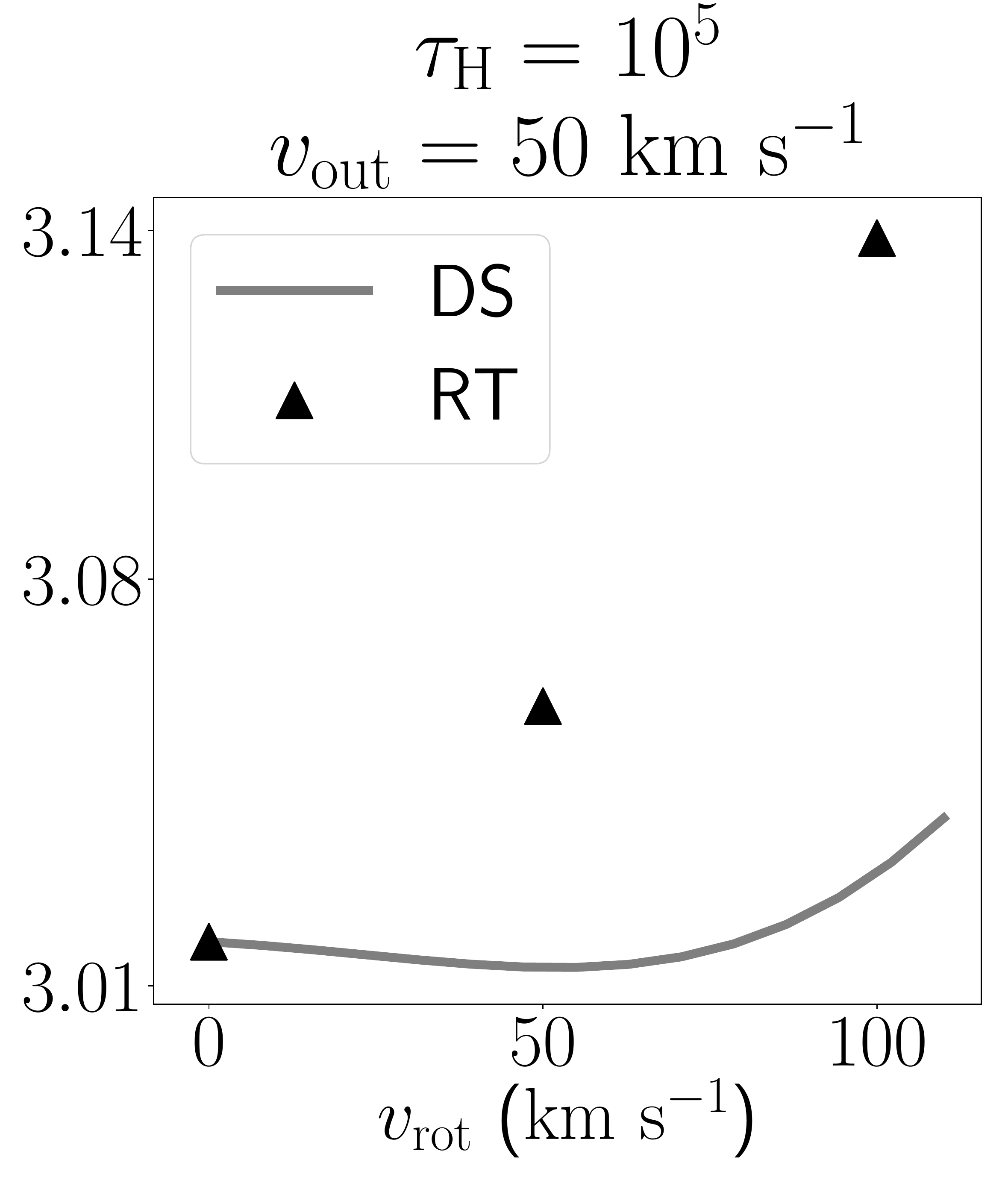}\\
\includegraphics[height=0.25\textwidth]{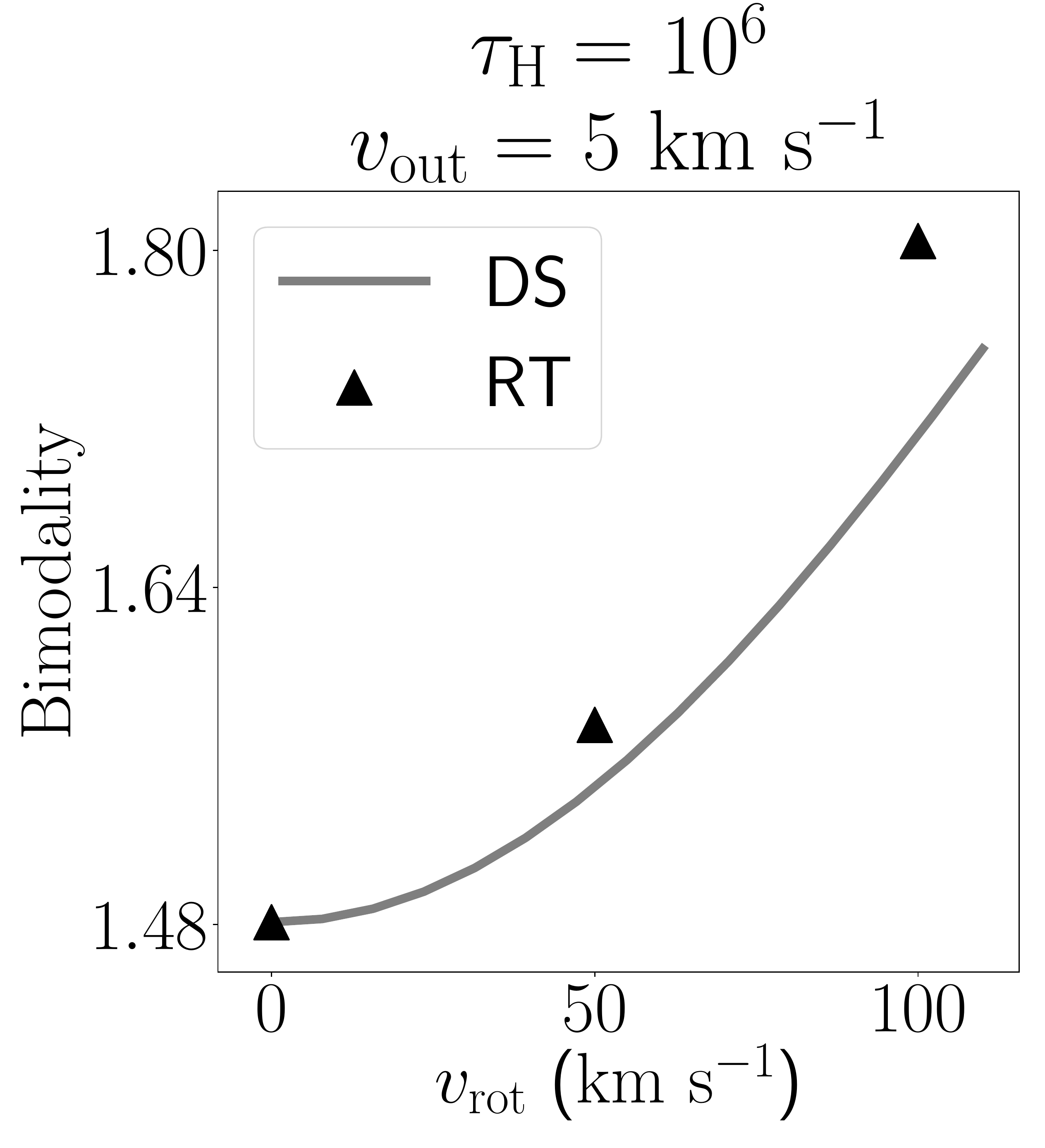}
\includegraphics[height=0.25\textwidth]{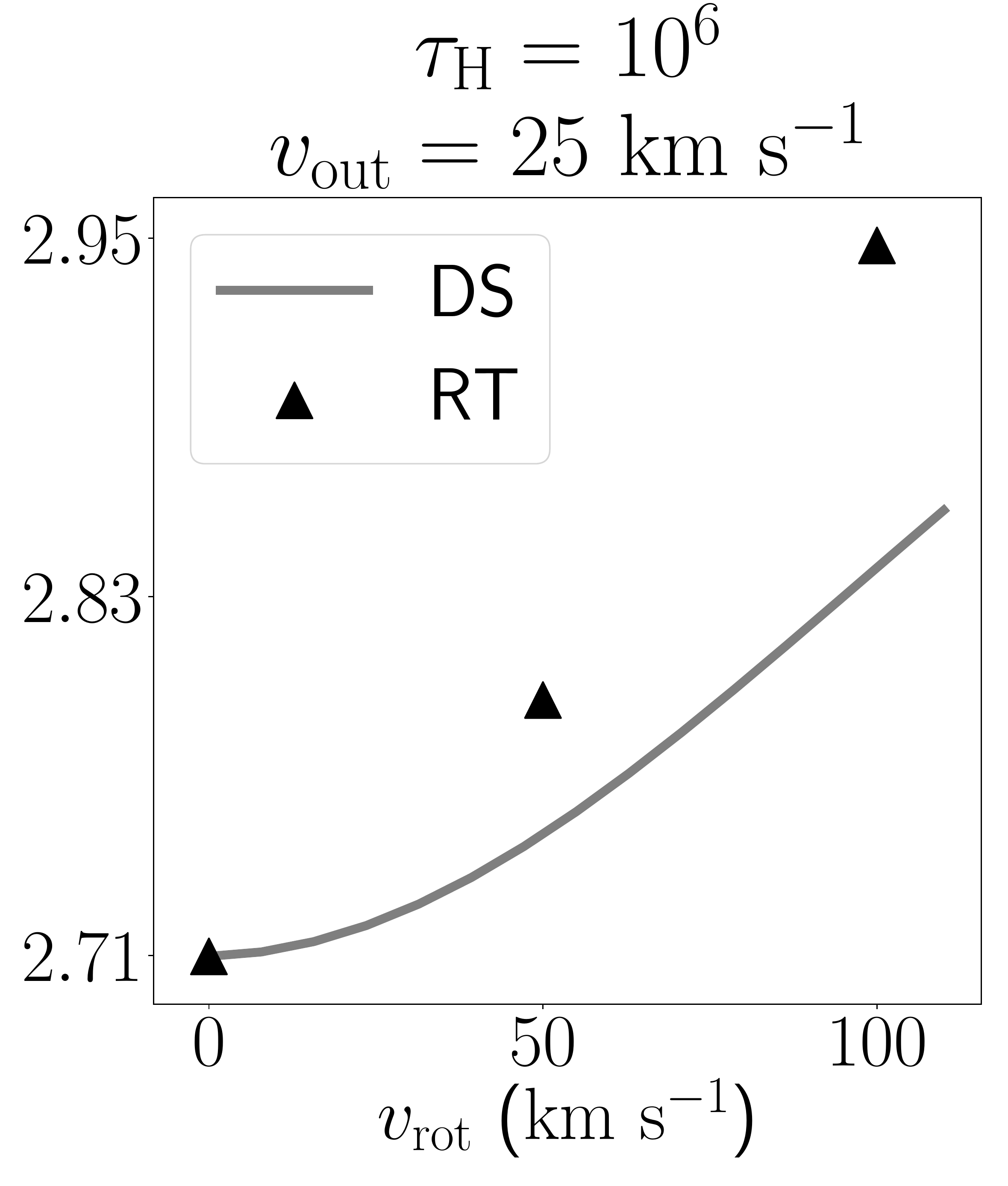}
\includegraphics[height=0.25\textwidth]{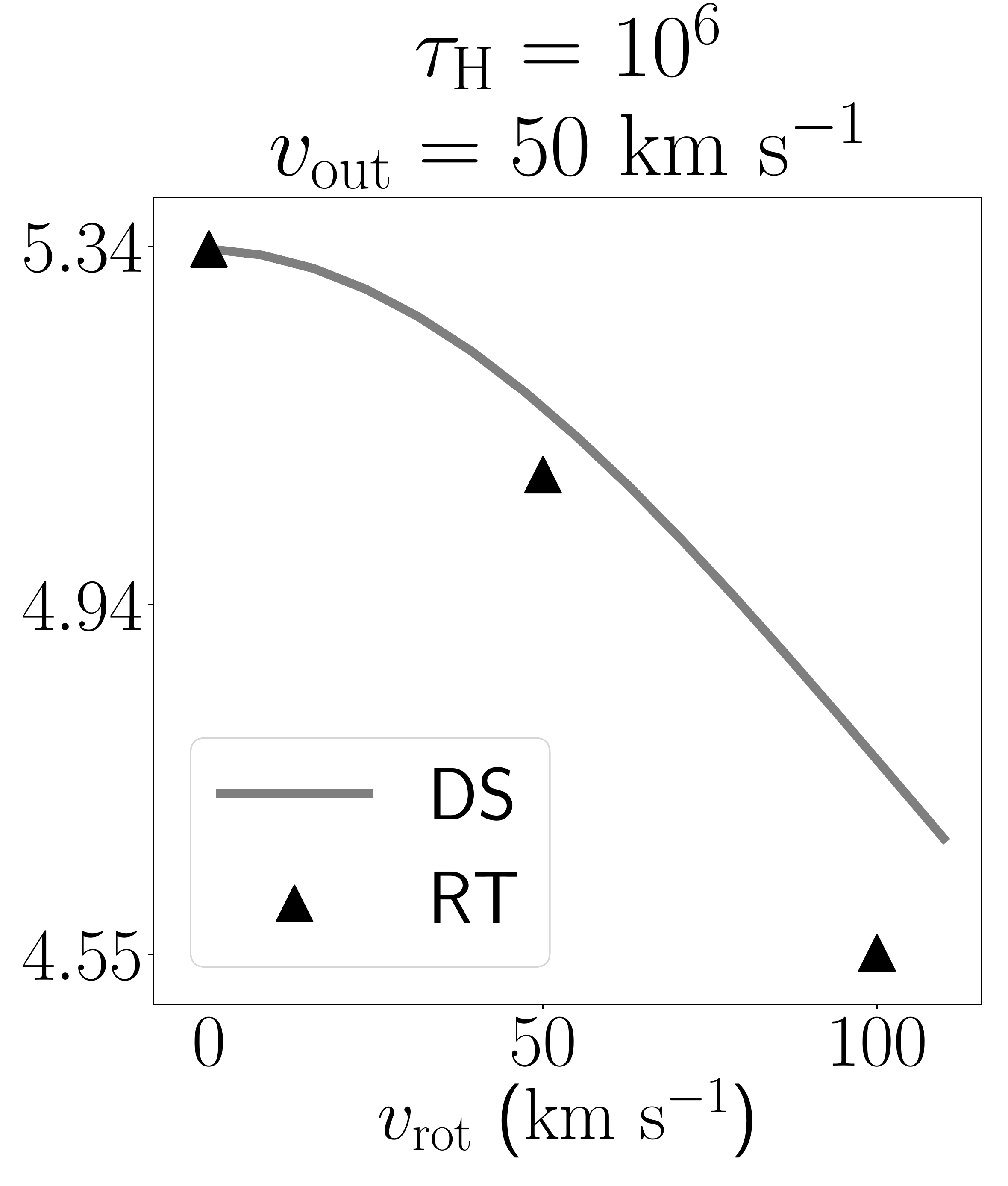}\\
\includegraphics[height=0.25\textwidth]{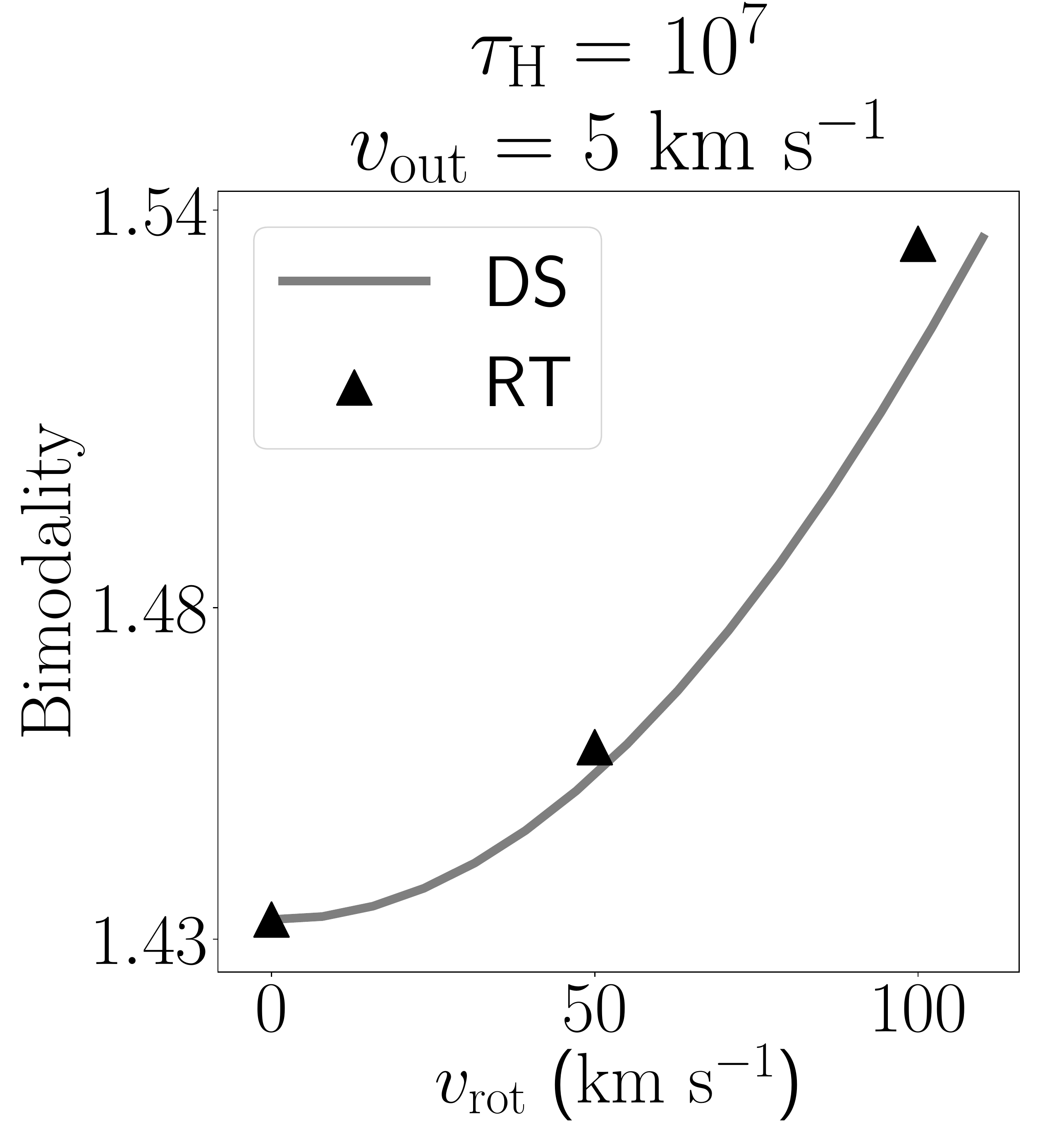}
\includegraphics[height=0.25\textwidth]{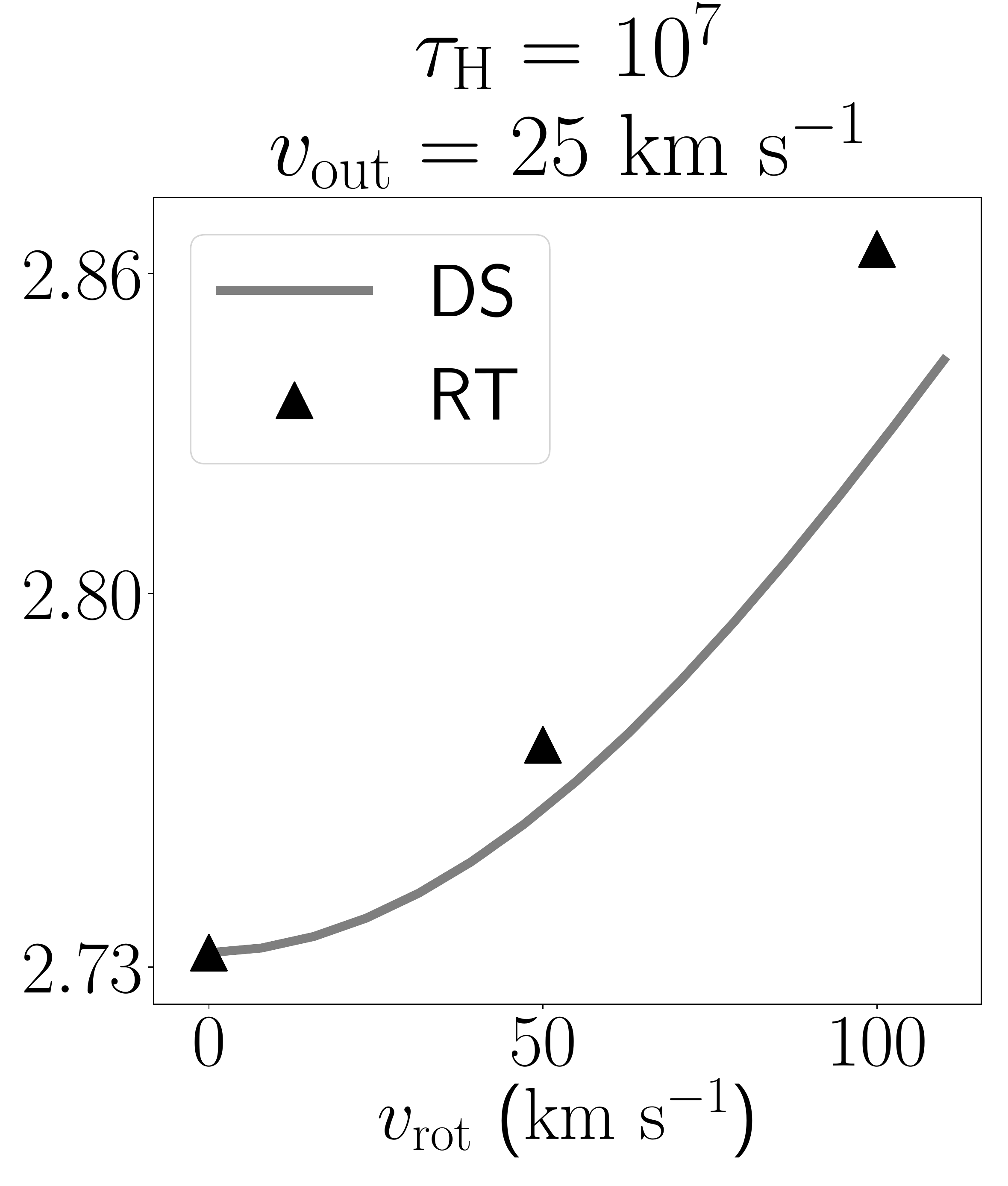}
\includegraphics[height=0.25\textwidth]{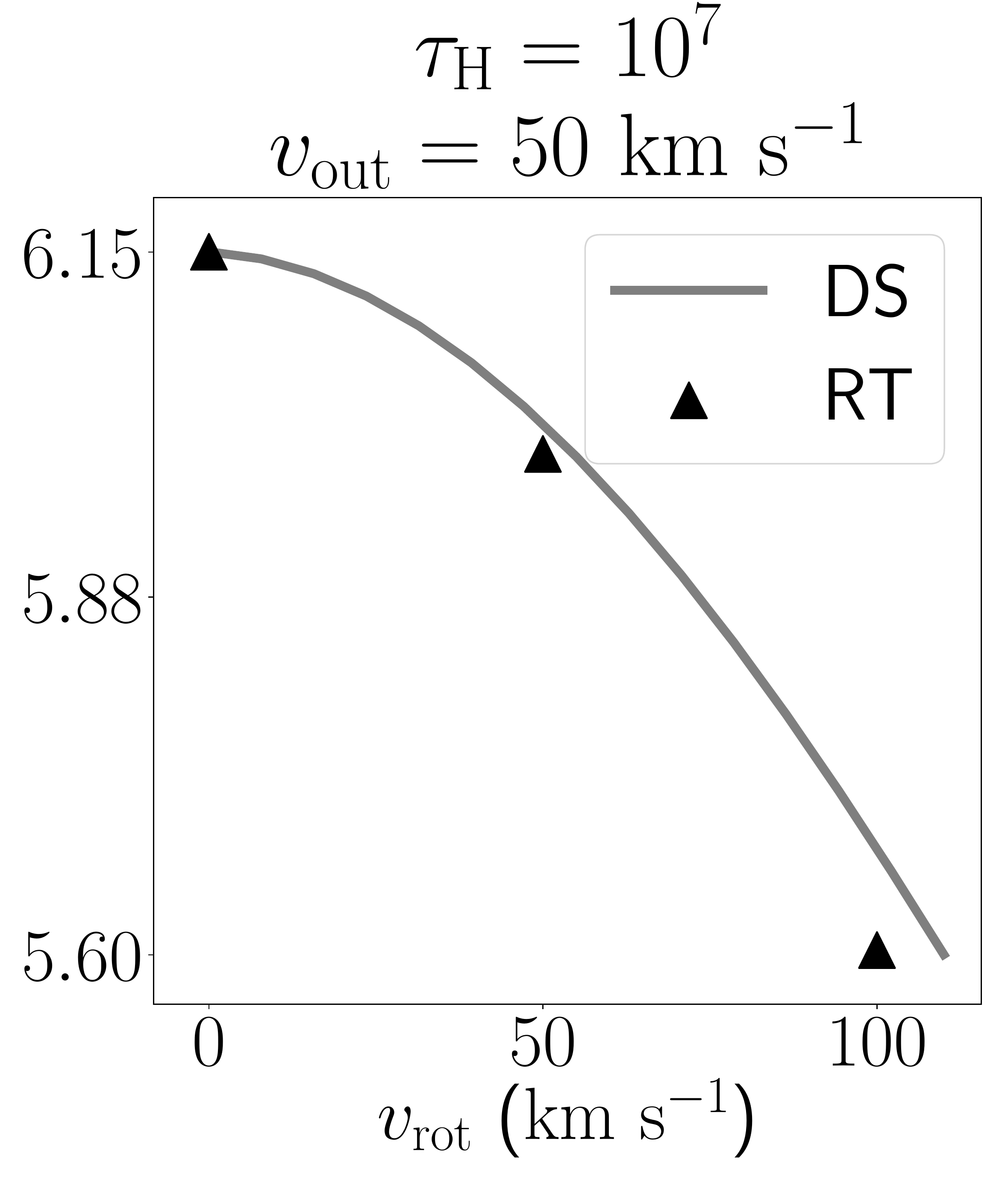}
\end{center}
\caption{\textbf{Bimodality trends.} Results for all the
  Radiative Transfer simulations (in triangles) compares against the
  Doppler Shift model (lines). 
  Follows the same layout as Figure \ref{fig:standard_deviation}. 
  \label{fig:bimodality}}
\end{figure*}

\begin{figure*}
\begin{center}
\includegraphics[height=0.25\textwidth]{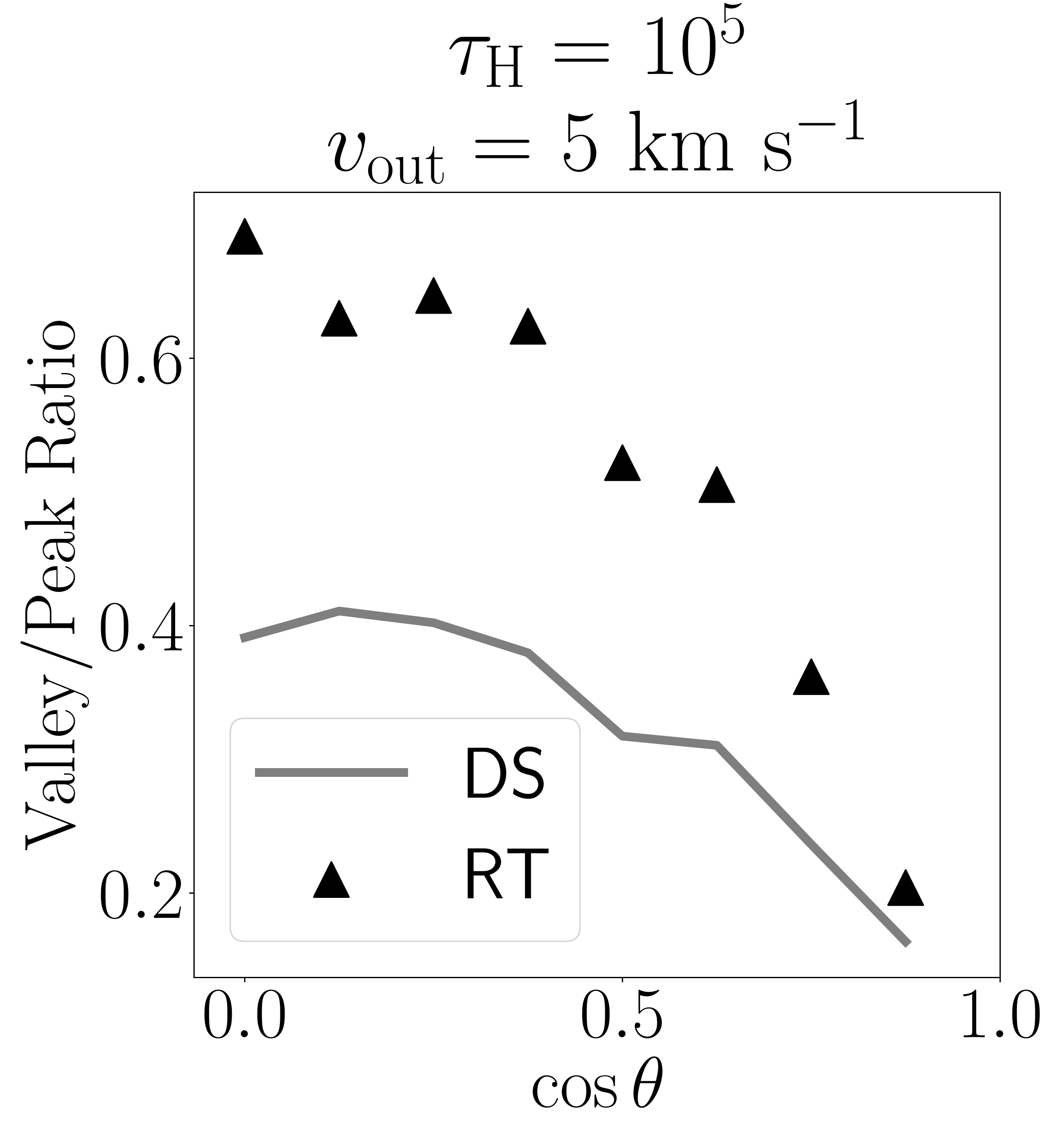}
\includegraphics[height=0.25\textwidth]{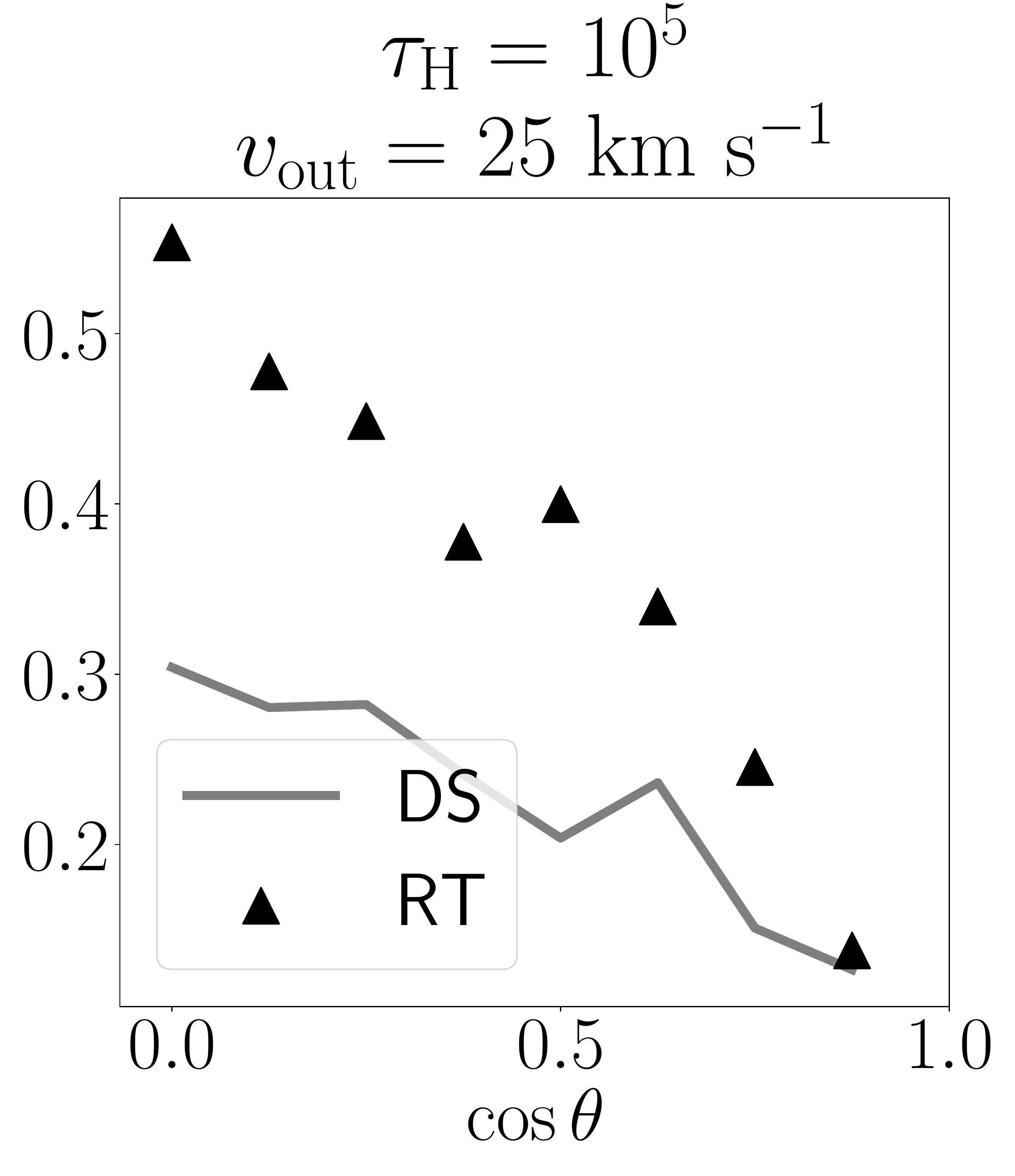}
\includegraphics[height=0.25\textwidth]{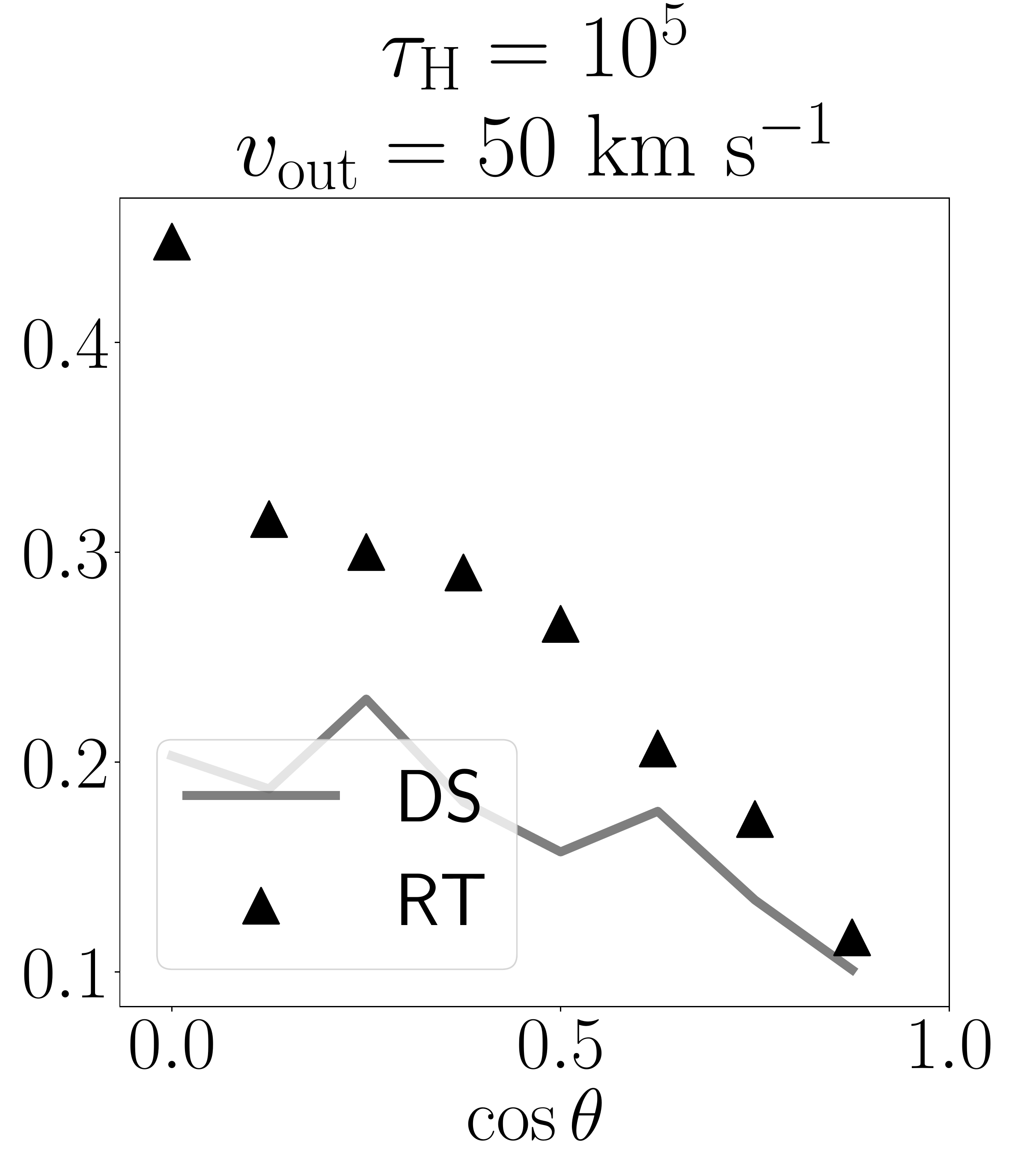}\\
\includegraphics[height=0.25\textwidth]{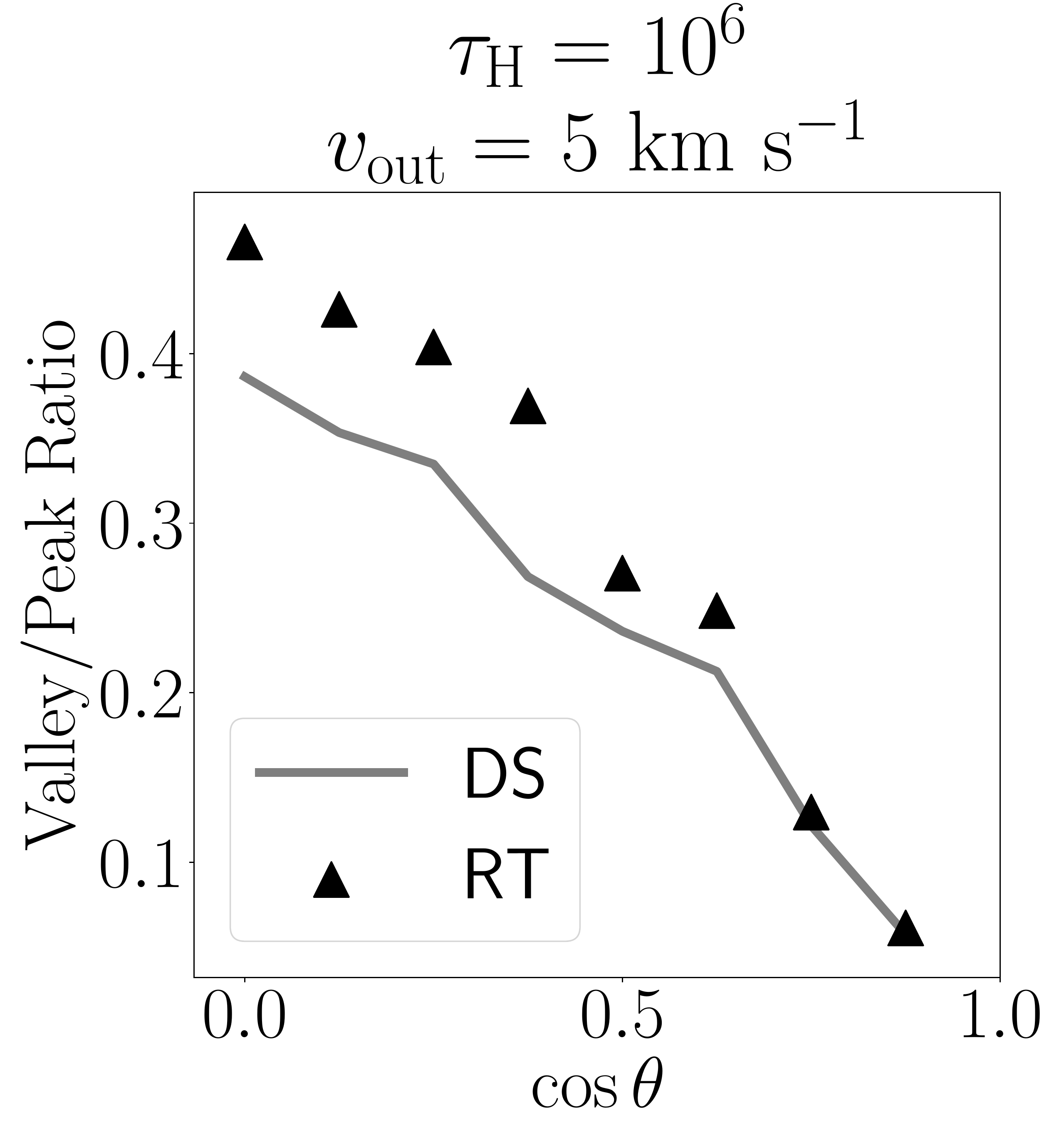}
\includegraphics[height=0.25\textwidth]{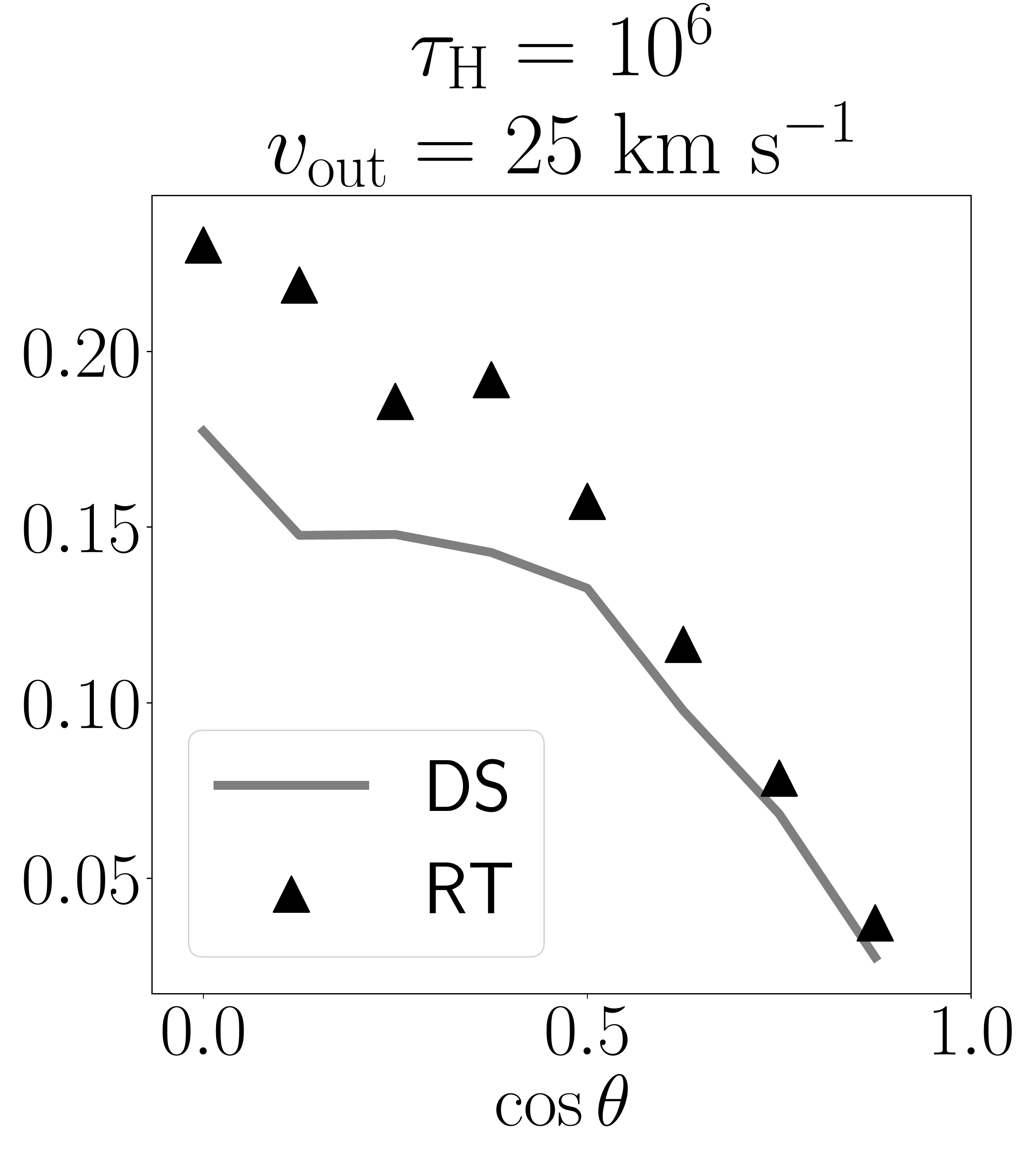}
\includegraphics[height=0.25\textwidth]{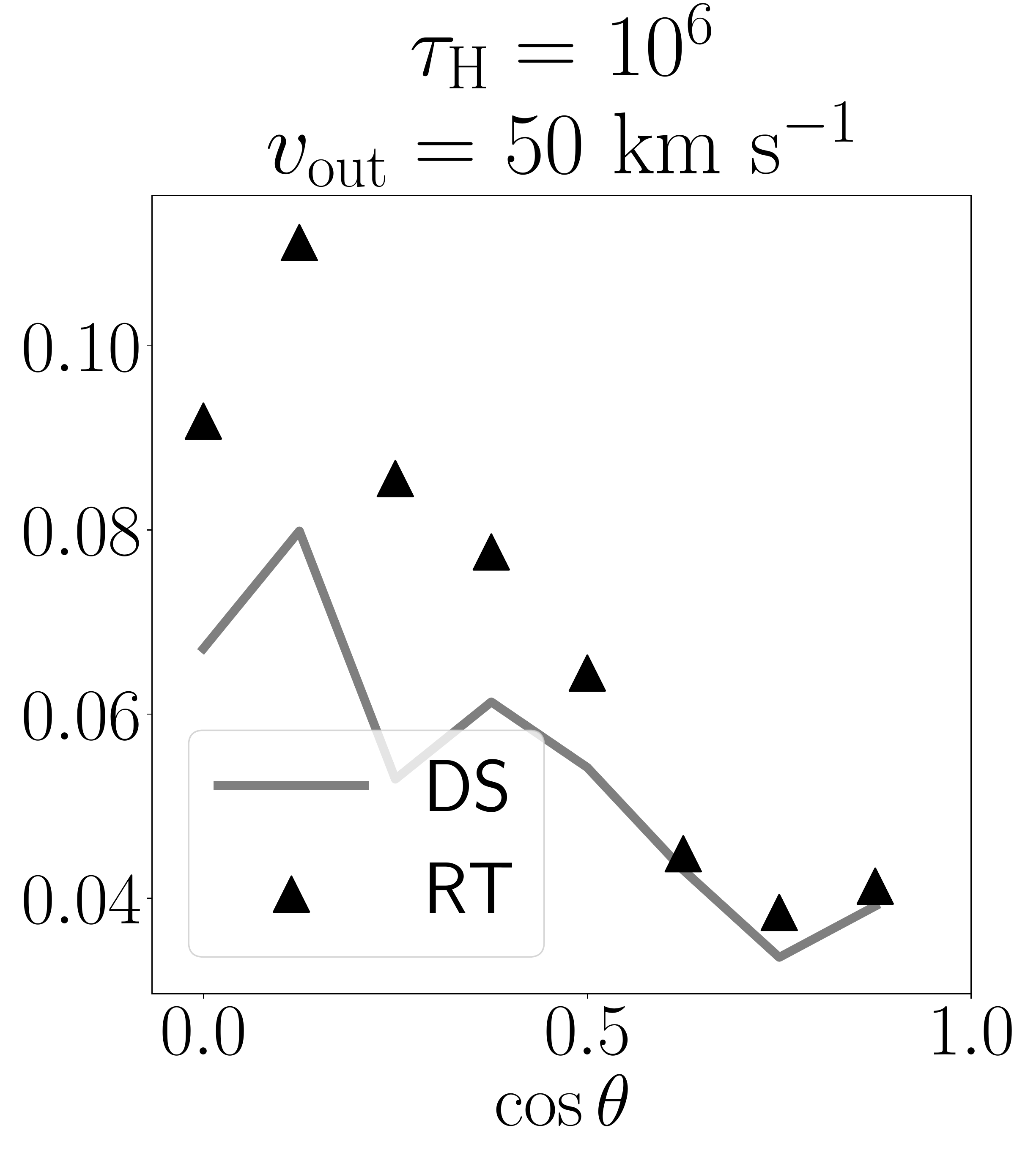}\\
\includegraphics[height=0.25\textwidth]{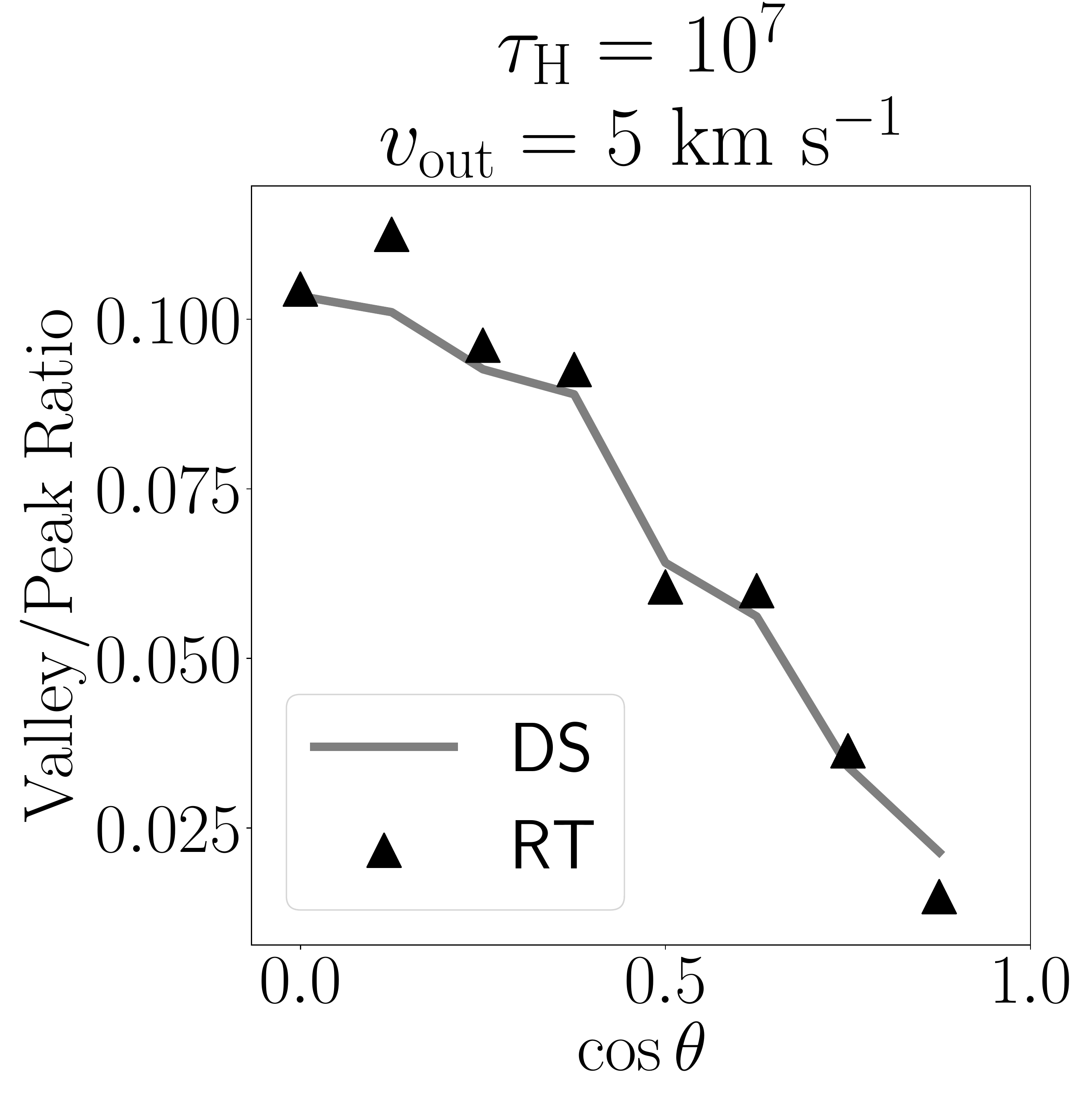}
\includegraphics[height=0.25\textwidth]{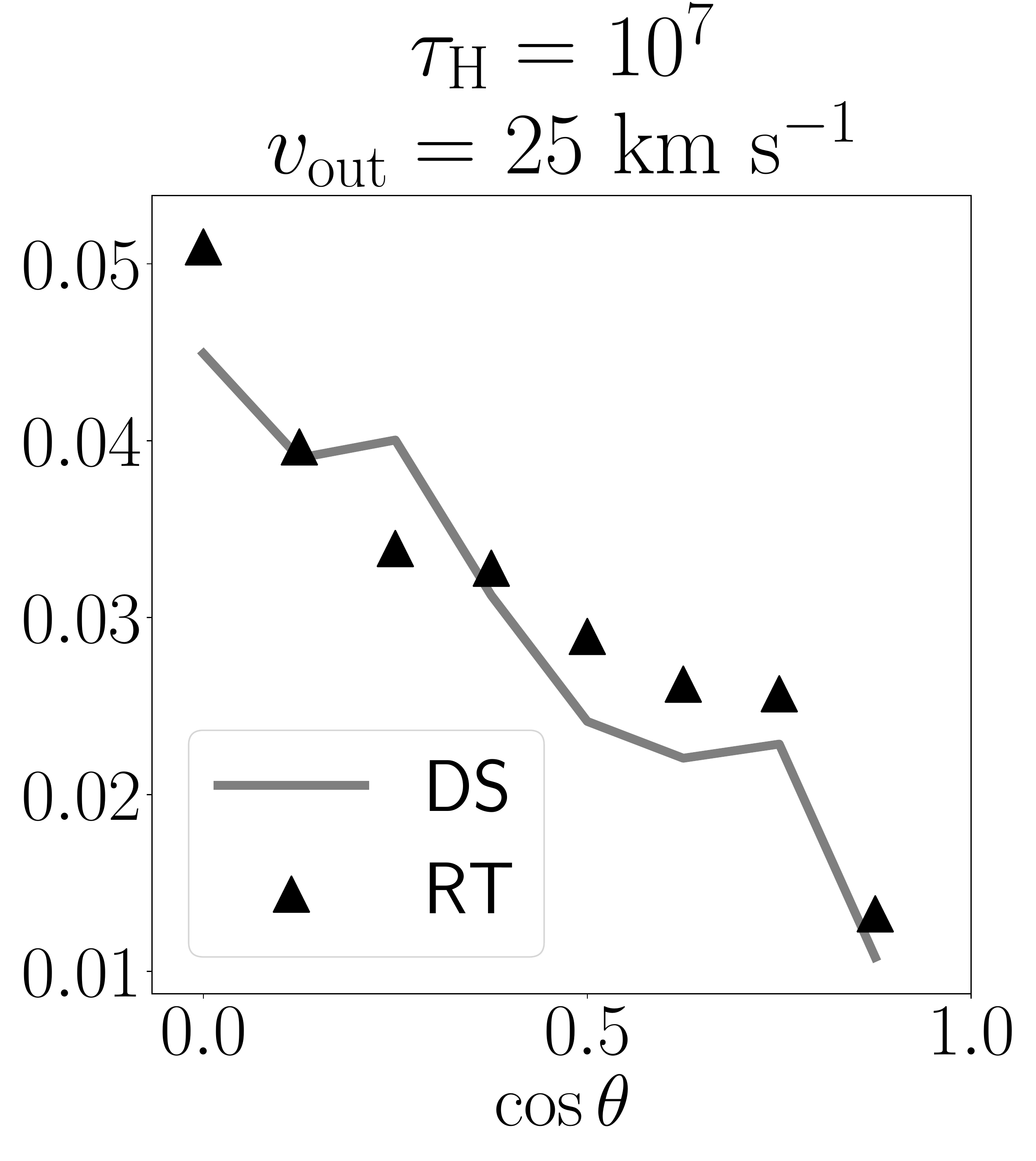}
\includegraphics[height=0.25\textwidth]{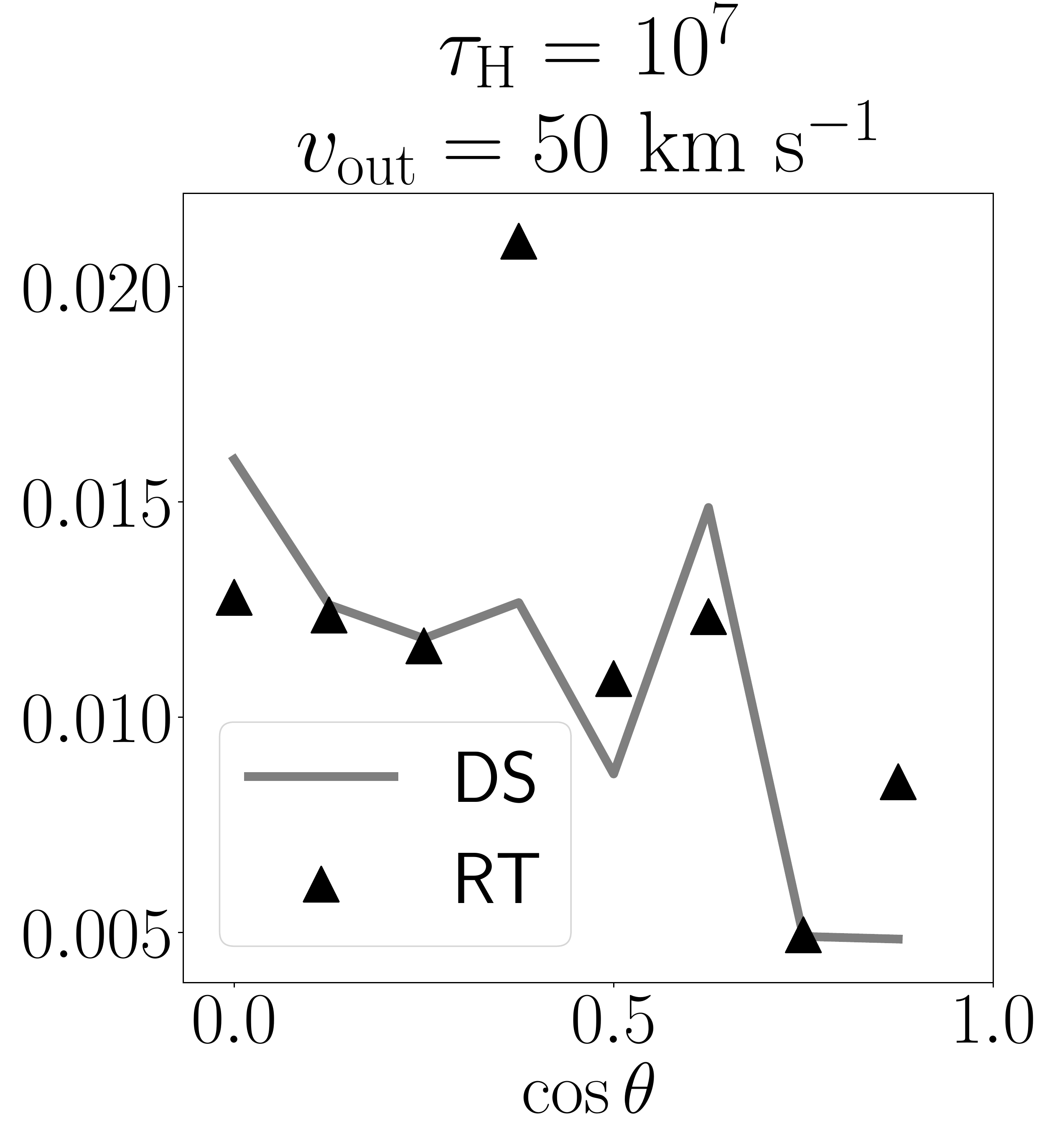}
\end{center}
\caption{\textbf{Valley Intensity. } We show for each \tauh the dependency that
  the viewing angle $\theta$ has on the line's the
  valley intensity. $\vrot=100$\kms is fixed for all panels.
		\label{fig:valley_intensity}}
\end{figure*}

\begin{figure}
\centering
    \includegraphics[width=0.48\textwidth]{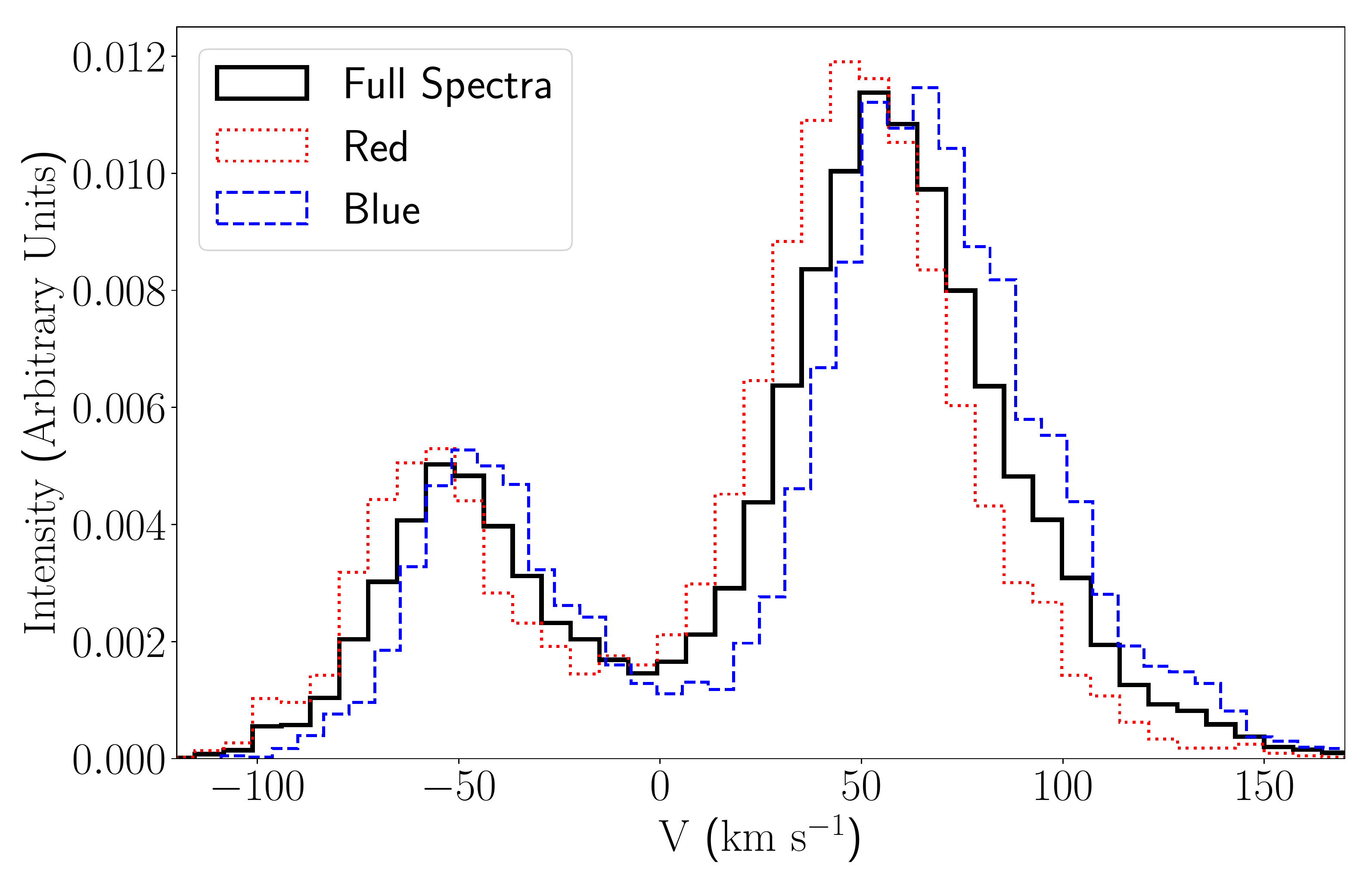}
  \caption{\textbf{Spectra from receding/approaching sides of a toy
      model LAE.}  These results correspond to the RT simulation with
    $\vout=25$\kms, $\vrot=50$\kms, $\tauh=10^5$. 
    The spectra were computed for a viewing angle of
    $\theta=90^{\circ}$. This toy model illustrates to what extent spectra from
    opposite sides of a galaxy have an imprint of the rotational
    kinematics. 
    \label{fig:doppler}}
\end{figure}

\section{Theoretical Models}
\label{sec:theory}

We use CLARA \citep{CLARA}, a Monte Carlo code that follows the
propagation of individual photons through a neutral Hydrogen medium
characterized by its temperature, velocity field and global optical depth.
The code assumes an homogeneous density throughout the simulated volume.
In the current implementation we neglect the influence of dust.
Our basic model is an spherical distribution of neutral hydrogen,
an approximation commonly used in the literature that explains a
wide variety of observational features
\citep{Ahn03,Verhamme06,Dijkstra06}. 

We use a velocity field that captures both outflows and rotation.
Outflows are described by a Hubble-like radial velocity profile with
the speed increasing linearly with the radial coordinate; this model
is fully characterized by $v_{\rm out}$, the radial velocity at the
sphere's surface. 
Rotation follows a solid body rotation profile fully characterized by
$v_{\rm rot}$ the linear velocity at the sphere's surface.  

The total velocity field is the superposition of rotation and
outflows.
The cartesian components take the following form:

\begin{equation}
	v_{x}=\frac{x}{R}\vout - \frac{y}{R}\vrot ,
	\label{eq:vx}
\end{equation}

\begin{equation}
	v_{y}=\frac{y}{R}\vout + \frac{x}{R}\vrot ,
	\label{eq:vy}
\end{equation}

\begin{equation}
	v_{z}=\frac{z}{R}\vout,
	\label{eq:vz}
\end{equation}
where $x$, $y$ and $z$ are the cartesian position coordinates with the
origin at the sphere's center, $R$ is the radius of the sphere and the
direction of the angular velocity vector is the $z$ axis.

For each run we follow $10^5$ individual photons generated from the
origin at the \lya line's center as they propagate
through the volume and finally escape.
We store the frequency and propagation direction for each photon
at its last scattering.

As input parameters we use $\tauh=\{10^5, 10^6, 10^7\}$,
$\vout=\{5,25,50\}\;\kms$ and $\vrot=\{0,50,100\}\;\kms$, for a total of
$27$ models with all the possible parameter combinations.
The values for the outflow velocity are lower than values commonly used in the
literature to allow for an interplay between the two kinematic features.
This range of parameters also produce emission lines with standard deviation and
skewness in the same range to those of high redshift LAES in recent
observations \citep{2017A&A...606A..12H}.

We also define the viewing angle, $\theta$, as the angle
between the rotation axis and the line of sight of a potential
observer. 

\cite{Garavito14} presented an analytical model that
accounts for the effects of pure rotation on the
\lya line morphology. 
The basic assumption of the model is that each differential surface
element on the sphere Doppler shifts (DS) the photons that it emits.
In this paper we introduce this ansatz by post-processing the results
of the outflows simulations without rotation.
The frequency of each photon is Doppler shifted as follows

\begin{equation}
x' = x + \frac{\vec{v}_{\rm rot}\cdot \hatk}{\vth}
\label{eq:shift_x}
\end{equation}
where $x'$ is the photon's new adimensional frequency, $x$ is the photon's
frequency after being processed only by the outflow, \vrot is the 
rotational velocity at the point of escape of the photon, \hatk is
the photon's direction of propagation and \vth is the thermal
velocity of the sphere. 
We fix \vth throughout the paper. This factor is a multiplying
constant that could be changed without running new Monte-Carlo
simulations, as the code works with the adimensional frequency $x$. 

This semi-analytic model allows us to produce new \lya spectra from
the outflow-only results and compare them with the full radiative
transfer solution including both outflows and rotation. 

\section{Results}
\label{sec:results}

\subsection{Qualitative Trends}
\label{sec:qualitative}

Figure \ref{fig:doppler_shift} summarizes the most important trends
from the RT simulations.
In the left side, the six panels correspond to $\tau=10^6$ and a viewing angle of
$\theta =90^{\circ}$, that is, perpendicular to the rotation axis of the
galaxy. 
In every panel the thin black line corresponds to the pure outflow
solution, i.e. without rotation. 
From top to bottom we see the effect of increasing the outflow
velocity, which is the expected increasing asymmetry towards the red
peak. 

The thick black line corresponds to the solution that includes both
outflows and rotation.
Comparing the left and right columns (lower versus higher rotational
velocity) we can see two immediate effects.
First, the line broadens and second, the intensity at the line's
center increases.

The thick gray line corresponds to the pure outflow solution
with the Doppler shift added to model rotation's influence.
At $\tauh=10^6$ the Doppler shift does a good job at capturing the broad
morphological features introduced by rotation: the angle dependence,
the broadening and the intensity increase at the line's center.

In the right side of Figure \ref{fig:doppler_shift} we show the same results as in the left one, but for a viewing angle of $\theta =
0^{\circ}$, that is parallel to the rotation axis. 
In this case we confirm the result presented by \cite{Garavito14},
namely that pure rotation introduces a strong dependence with 
viewing angle, a trend that we find also holds for rotation mixed with
outflows.   

The quality of the results from the Doppler shift improves for higher
\tauh values. 
In the Appendix we show the same plots as Figure
\ref{fig:doppler_shift}, there it is evident that for $\tauh=10^5$ the
results are not as good as they are for $\tauh=10^6$, and that for
$\tauh=10^7$ the Doppler shift provides a remarkable good
approximation. 

\subsection{Quantitative Trends}
\label{sec:quantitative}

After finding the qualitative influence of the different parameters we
move onto a quantitative study.
To do this we summarize the line morphology by four different
scalars: standard deviation (STD), skewness (SKW), bimodality
(BI) and valley/peak ratio.
These quantities are defined by the following equations \citep{kokoska1999}:

\begin{equation}
\label{eq:std}
\STD = \sqrt{m_2},
\end{equation}

\begin{equation}
\label{eq:skw}
\SKW = \frac{m_3}{m_2^{3/2}},
\end{equation}

\begin{equation}
\label{eq:bi}
\BI = \mathrm{KURTOSIS} - \SKW^2 = \frac{m_4}{m_2^{2}} - \frac{m_3^2}{m_2^{3}},
\end{equation}
where each $m_i$ is the i-th moment about the mean. 
The STD has velocity units and quantifies the line's width.
The SKW is adimensional and quantifies the peaks' asymmetry. 
In the case of a bimodal distribution, $\SKW>0$ means that the blue
peak is taller and for $\SKW<0$ the red peak is taller. 
The BI is adimensional and quantifies whether the line has 1 or 2
peaks: it is  always $\geq 1$ \citep{Pearson1929} and the closer 
to 1, the more bimodal is the line (i.e. has 2 similar peaks). 
We found by visual inspection of our spectra that $\BI=2.5$ marks the
transition between two peaks (however imbalanced) and a dominant
single peak.

\subsubsection{Standard Deviation}
Figure \ref{fig:standard_deviation} summarizes the standard deviation
results for all our models.
Each panel shows the STD as a function of \vrot.
All panels were computed using a viewing angle of $\theta =
90^{\circ}$ (perpendicular to the rotation axis), which has the most
extreme influence from rotation.
 The black triangles
correspond to the full RT solution and the line to the DS
approximation.  
The optical depth increases from top to bottom and the outflow
velocity from left to right.
This quantitative plot confirms that the line width increases with
rotational velocity and optical depth.
These trends are expected; higher rotational velocities can be seen as
an addition of different Doppler shifts that smear out the line, while
a higher optical depth translates into a larger number of scatterings
that increase the probability of a photon to diffuse in frequency
resulting in a broader line.

The DS successfully reproduces all trends with the optical depth,
rotational velocity and outflow velocity.
However, the DS consistently underestimates the STD. 
The difference between the RT and DS increases with the outflow
velocity and the rotational velocity, and decreases with increasing
optical depth.
In the range of parameter space explore, this difference has as an
upper bound of $\sim 7\%$, $3\%$ and $\sim 2\%$ for
 $\tauh=10^5$, $10^6$ and $10^7$, respectively. 

\subsubsection{Skewness}

Figure \ref{fig:skewness} presents the skewness results for all the
models together with the DS comparison following the same layout as
Figure \ref{fig:standard_deviation}.
In all cases the skewness is negative showing that all the lines
are unbalanced towards the red side of the spectrum.
Skewness increases with rotational velocity and decreases with
optical depth; rotation tries to smooth the line diminishing the
asymmetries while a higher optical depth reinforces the line asymmetries.
The skewness does not have a monotonous trend with outflow velocity because
there is a transition between double and single peak line; for low
outflow velocities the skewness signals the balance between the two
existing peaks while for high outflow velocities it quantifies the
asymmetry of the already dominant read peak.

The DS reproduces the main trends, again with an underestimation that
decreases at higher optical depths and increases with larger values of
the rotational velocity and outflow velocity.
In this case the differences between RT and DS have an upper bound of
$85\%$, $35\%$ and $5\%$ 
for  $\tauh=10^5$, $10^6$ and $10^7$,
respectively.

\subsubsection{Bimodality}

Figure \ref{fig:bimodality} shows the results for the bimodality using
the same layout as in the two previous Figures.
Following the reasoning about the skewness, we observe that
increasing the outflow velocity increases the value of bimodality,
that is, it transitions to a more pronounced single peak. 
The trend as a function of the rotational velocity and the optical
depth are not monotonous.
When the outflow velocity is low ($\vout<50\ \kms$), an increasing
rotational velocity smears the two asymmetrical peaks pushing the line
morphology towards a single peaks, making the bimodality statistics
increase. 
On other situations ($\vout=50\ \kms$ and $\tauh\geq 10^6$) higher
rotational velocities the bimodality statistics decreases, which means
that it manages to slightly enlarge the already dominant red peak.

The DS reproduces the main trends while underestimating the bimodality
statistics. 
As expected from the previous results the difference between RT and DS
decreases at higher optical depths and increases with increasing
values of the rotational and outflow velocities.
In this case the differences have an upper bound of $4\%$, $2\%$ and
$1\%$ for  $\tauh=10^5$, $10^6$ and $10^7$, respectively.  

\subsubsection{Intensity at line's center}

In Figure \ref{fig:valley_intensity} we quantify how the intensity at
the line's center (i.e. the valley) changes with the viewing angle,
the outflow velocity and the optical depth.
These results correspond to a fixed rotational velocity of
$\vrot=100\ \kms$.
The triangles correspond to the RT simulations and the line represents
the DS results. 
The valley intensity is expressed as a fraction of the maximum peak
intensity in the line, as such the valley/peak ratio is always $<1$. 
In every panel we see that the valley/peak ratio decreases as the
observer moves from a line  of sight perpendicular to the rotation
axis onto a parallel line of sight. 
This is a clear demonstration of the viewing angle dependency
introduced by rotation.

The valley/peak ratio at $\cos{\theta}=1$ matches results
without rotation, this shows that for increasing rotational velocity
the valley/peak ratio increases.
In turn, for increasing optical depth or outflowing velocity this
ratio decreases.
Once again, the DS results correctly follow the trends for the full RT
simulations.  
This time the differences have an upper bound of $55\%$, $2\%$ and
$1\%$ for  $\tauh=10^5$, $10^6$ and $10^7$, respectively.

\section{Discussion}
\label{sec:discussion}

In this section we discuss how the results we have presented can be
connected to the interpretation of observational data.

\subsection{The semi-analytic model as gaussian smoothing}

The effects of the semi-analytic model on the pure outflowing spectra
are similar to the expected results from a gaussian smoothing, 
such smoothing can also be a natural consequence of the intruments
used to measure the LAE spectra.
As a consequence, natural experimental artifacts could be mistaken as
an indication for rotation.

To understand to what extent the effects of rotation can be seen as a
simple smoothing, we model the effects of gaussian smoothing with a
similar ansatz as  the semi-analytic model by changing the frequency
of each photon by a Gaussian random variable centered on zero and
standard deviation $\sigma_x$: \begin{equation}
x^{\prime} = x + \mathcal{N}(0,\sigma_x^2).
\end{equation}

We find that the results of the semi-analytic model on the total
spectra at a given angle $\theta$ can be approximated using
$\sigma_x = (1/2)\times(\vrot/\vth)\times \sin\theta$. 
This approximation reproduces within a few porcent all quantities
measured in Section \ref{sec:quantitative} for the semi-analytic
model, it only works well for $\tau=10^6$ and $\tau=10^7$. 

This result can be used as an order of magnitude estimate of the mimum
espectral resolution required to observe the effects of a rotational
velocity of $\vrot$. 
The Space Telescope Imaging Spectrograph (STIS) on the Hubble Space
Telescope (HST) can provide resolutions around 20 \kms
\citep{2015PASA...32...27H}, which could allow for the detection of
rotation features in nearby galaxies with at least $40$ \kms and a
rotation axis perpendicular to the line of sight.  
Other instruments such as the Cosmic Origins Spectrograph (COS) have
varying resolutions between $20$\kms and $200$\kms according to the
angular size of the source \citep{2018A&A...616A..60O}.

\subsection{\lya Kinematic Maps}
\label{subsec:kinematic}

Current observational facilities have the capability of spatially
resolving the extent of a LAE.
For instance \cite{Prescott14} presented observational results of a
Doppler shift when taking spectra at two opposite sides of a large
($\approx 80$kpc) LAE.
In more recent work \cite{2018MNRAS.473.3907A} mapped \lya emission
around a quasar.
In this case one could measure the displacement between the
peaks in the two different spectra. 
The interest of this test is that the location of the peaks only
depends on the kinematic properties of the sources and does not
change by convolution with a gaussian window function due to limited
instrumental resolution, as discussed in the previous section.

In Figure \ref{fig:doppler} we present a toy model (\vrot$=50\ \kms$,
\vout$=25\ \kms$ and \tauh$=10^5$) for the spectrum of
a LAE taken from to different sides of the galaxy. 
As the LAE is rotating, one side is being redshifted while the other
is blueshifted. 
We see that the full spectrum is a weighted line, in solid black,
that is found between these two.
We notice that the distance between the maxima of the blue and red
spectra is not twice the rotational velocity as it could be naively
expected.

In this toy model the distance between the peaks of the
receding/approaching spectra is close to $\sim 25\ \kms$, which is
a fourth of the naively expected value of $2\vrot=100\ \kms$,
due to the fact that only a small fraction of the photons are emitted
at the extreme of the galaxy having the maximum rotational velocity of
$50\ \kms$. 
Although it is a good approximation to think the rotating spectra by a
sum of Doppler shifts, the peak of the spectra is also weighted by the
amount of mass with a given line-of-sight velocity. 

Spectrographs like the Multi Unit Spectroscopic Explorer (MUSE) could
obtain kinematic information from large samples of LAEs to build
velocity maps in Ly$\alpha$.  
This could be a natural extension of the work reported by
\cite{Herenz2016} on the velocity maps of several LARS (Lyman Alpha
Reference Sample) galaxies.  
The interpretation of such data could take into account the insights
and trends we have presented in this paper.

\section{Conclusions}
\label{sec:conclusions}

In this paper we explore, for the first time in the literature,
the results of a model for the emergent \lya line from rotating
outflows. 
We use a semi-analytic model first presented by \cite{Garavito14}
to capture the main effects of rotation and confront it against
results from Monte-Carlo radiative transfer simulations. 
The semi-analytic model only takes into account the Doppler shift computed
as product of quantities at the surface of last scattering, namely
$\vec{v}_{\rm rot}\cdot \hatk$, where $\vec{v}_{\rm rot}$ is the
velocity due to rotation and \hatk is the direction of the photon's
propagation. 

To address the first question we posed in the introduction (\emph{What
  is the expected imprint of rotation on a resonant emission line such
  as the \lya line?}) we find from the full RT simulations that the
effects of rotation on the \lya line morphology are:  

\begin{itemize}
  \item Inducing a dependency on the viewing angle.
  \item Broadening the line.
  \item Increasing the intensity at the line's center.
\end{itemize}
All these effects can be qualitatively explained by the proposed
semi-analytic model. 
Quantitatively speaking the semi-analytic model provides a
satisfactory answer for a neutral Hydrogen optical depth equal or
larger than $10^6$.

Addressing the second question we posed in the introduction (\emph{To what
extent is it possible to constrain rotational kinematics from the \lya
emission line?})  we show that the most straightforward approach
is measuring \lya spectra from two different regions of a
galaxy to detect approaching/receding gas motions
\citep{Prescott14,2018MNRAS.473.3907A}. 
In that case, we also show that the distances between these two peaks 
is four times shorter than the naively expected value of
$2\times\vrot$.  
This difference is produced by the different geometric weights at the
emitting surface.    
Here the semi-analytic model also provides the correct quantitative
insight. 

To summarize, our works shows that the Doppler Shift offers an
easy-to-implement approximation to explore such influence into already
existing radiative transfer simulations that include
outflows, providing a versatile tool to interpret current and future \lya
kinematics maps \citep[e.g][]{2018MNRAS.473.3907A,Erb18}.   

\bibliographystyle{mnras}
\bibliography{references}

\appendix

\section{Additional figures}
\label{sec:appendix}

\begin{figure*}
  \begin{center}
    \includegraphics[width=0.49\textwidth]{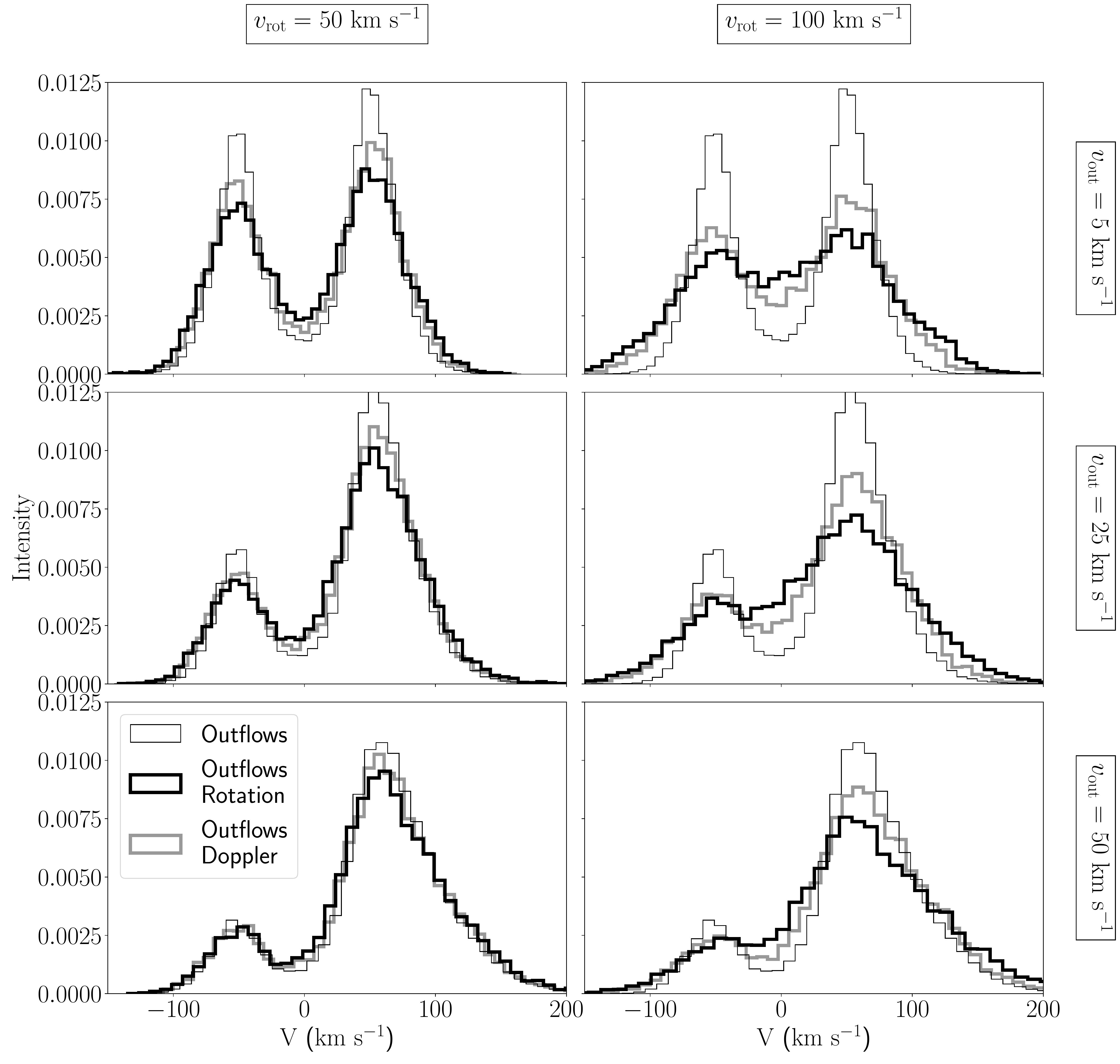}
    \includegraphics[width=0.49\textwidth]{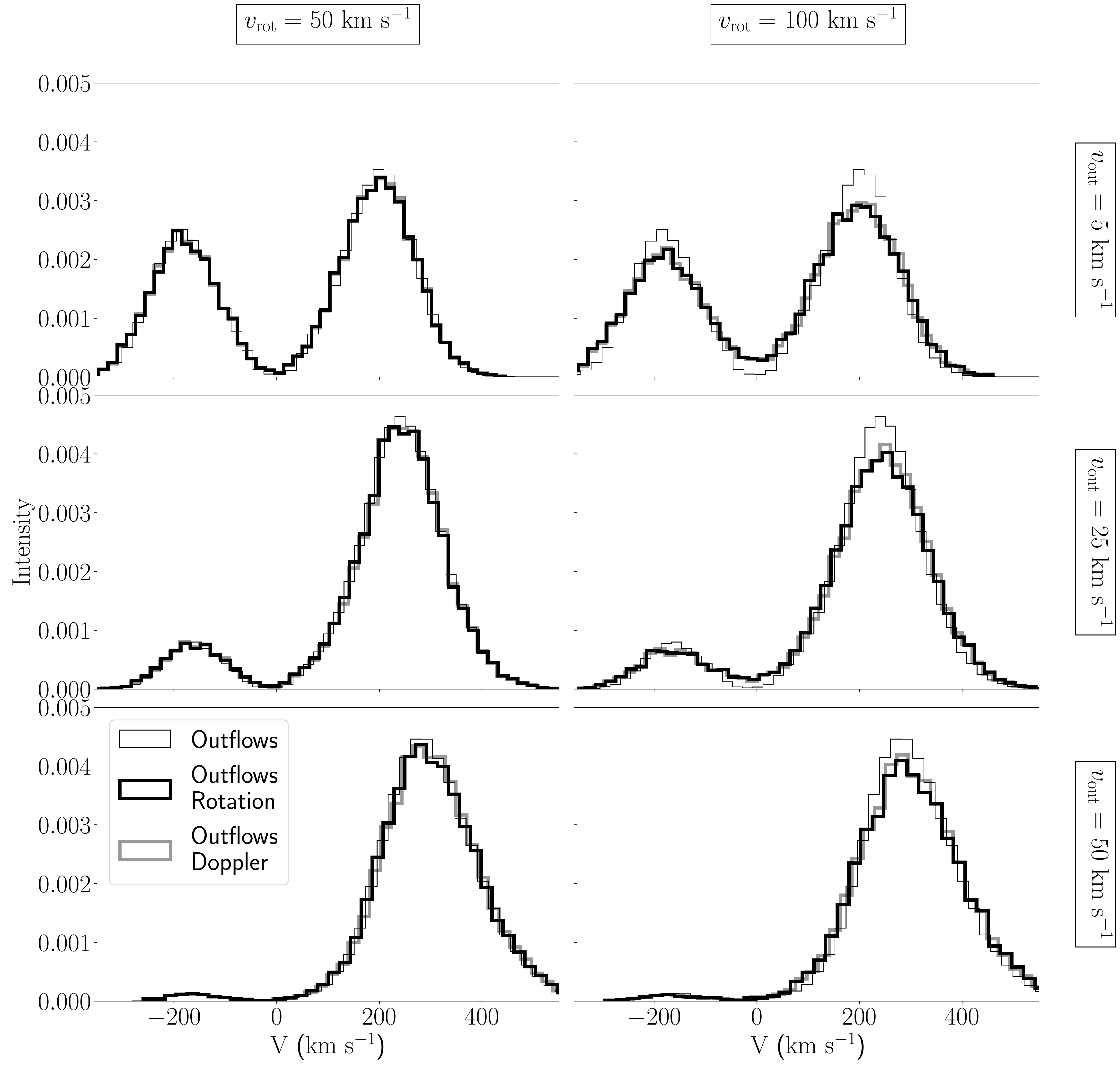}
  \end{center}
  \caption{\textbf{Qualitative trends of changing outflow and
      rotational velocity.}
    Same layout as Figure \ref{fig:doppler_shift}. On the left:
    $\tauh=10^5$ and $\theta=90^\circ$; on the right:     this time  $\tauh=10^7$ and $\theta=90^\circ$.}
\end{figure*}

\end{document}